\documentclass{osa-article}
    
\journal{oe}

\articletype{Research Article}

\usepackage[utf8]{inputenc}
\usepackage{xspace}
\usepackage{subcaption}

\newcommand{\norme}[1]{\left\Vert #1\right\Vert}
\newcommand{\abs}[1]{\left\vert #1\right\vert}
\def\ddroit{\mathrm{d}}

\DeclareFontFamily{U}{wncy}{}
\DeclareFontShape{U}{wncy}{m}{n}{<->wncyr10}{}
\DeclareSymbolFont{mcy}{U}{wncy}{m}{n}
\DeclareMathSymbol{\Sh}{\mathord}{mcy}{"58} 

\newcommand{\note}[1]{{\em\textbf{\color{red}[#1]}}}
\renewcommand{\note}[1]{} 

\newcommand{\notevf}[1]{{\em\textbf{\color{blue}{#1}}}} 
\renewcommand{\notevf}[1]{#1} 

\newcommand{\myref}[2]{\ref{#1}(\subref{#2})}

\begin{document}

\title{Jointly super-resolved and optically sectioned Bayesian reconstruction method for structured illumination microscopy}

\author{Yann Lai-Tim,\authormark{1,2,*} Laurent M. Mugnier,\authormark{1} François Orieux,\authormark{3} Roberto Baena-Gall\'e,\authormark{4,5} Michel Paques,\authormark{2} and Serge Meimon\authormark{1}}

\address{\authormark{1}DOTA, ONERA, Université Paris Saclay, F-91123 Palaiseau, France\\
\authormark{2}CIC 503, INSERM, Quinze-Vingts Hospital, Paris, France\\
\authormark{3}Laboratoire des Signaux et Systèmes (Univ. Paris-Sud, CNRS CentraleSupélec, Université Paris-Saclay), 3 rue Joliot-Curie, 91 192 Gif-sur-Yvette, France\\
\authormark{4}Instituto de Astrofísica de Canarias (IAC), 38205 La Laguna, Tenerife, Spain\\
\authormark{5}Departamento de Astrofísica, Universidad de La Laguna (ULL), 38206 La Laguna, Tenerife, Spain}

\email{\authormark{*}yann.lai-tim@onera.fr} 



\begin{abstract}
Structured Illumination Microscopy (SIM) is an imaging technique for achieving both super-resolution (SR) and optical sectioning (OS) in wide-field microscopy. It consists in illuminating the sample with periodic patterns at different orientations and positions. The resulting images are then processed to reconstruct the observed object with SR and/or OS.
In this work, we present BOSSA-SIM, a general-purpose SIM reconstruction method, applicable to moving objects such as encountered in \emph{in vivo} retinal imaging, that enables SR and OS jointly in a fully unsupervised Bayesian framework.
By modeling a 2-layer object composed of an in-focus layer and a defocused layer, we show that BOSSA-SIM is able to jointly reconstruct them so as to get a super-resolved and optically sectioned in-focus layer.
The achieved performance, assessed quantitatively by simulations for several noise levels, compares favorably with a state-of-the-art method. Finally, we validate our method on open-access experimental microscopy data.
\end{abstract}

\section{Introduction}

Structured Illumination Microscopy (SIM) is a commonly used imaging technique in wide-field fluorescence and reflectance microscopy for achieving optical sectioning (OS) \cite{neil_method_1997} and super-resolution (SR) \cite{gustafsson_surpassing_2000}. It consists in projecting usually periodic light patterns so as to spatially modulate the intensity response of the observed object. The induced modulation produces two effects in the acquired images.
On the one hand, it down-modulates high frequency object components that were previously inaccessible into the support of the optical transfer function (OTF), thus enabling super-resolved reconstruction. On the other hand, the modulated content of the images essentially comes from object layers that lie near the focal plane as the modulation contrast strongly decreases with defocus; this allows one to extract from the modulated frequency components an optical section of the object. 
By illuminating the sample for different spatial phases and orientations of the light pattern, these two attributes of SIM can be achieved, separately or jointly.

On the applicative side, recent works have tried to extend the scope of SIM to \textit{in vivo} imaging and in particular to retinal imaging~\cite{shroff_phaseshift_2009,shroff_lateral_2010,gruppetta_theoretical_2011,chetty_structured_2012}, for which the SR and OS brought about by structured illumination are very promising. The main issue that was discussed is the uncontrolled eye motion, which causes inter-frame motion. This is a problem for conventional SIM implementations~\cite{neil_method_1997,gustafsson_surpassing_2000}, which rely on accurately phase shifting the illumination patterns over a static sample.
Conversely, the object motion can be used to introduce phase shifts without having to translate the illumination patterns, although the phase shifts are uncontrolled in this case. Following this idea, 2D SIM reconstruction methods for moving objects have been proposed~\cite{shroff_lateral_2010,chetty_structured_2012}. They firstly use image registration so as to get images where the object is aligned and the illumination patterns shifted over the object. Then, the resulting phase-shifts are estimated~\cite{shroff_phaseshift_2009} and the reconstruction is conducted by solving a system of linear equations. However, these methods achieve SR~\cite{shroff_lateral_2010} or OS~\cite{chetty_structured_2012} independently and not jointly.

In this paper, we propose a general-purpose SIM reconstruction method applicable to moving objects that jointly performs OS and SR, with retinal imaging in mind. The novelty of our approach is twofold. Firstly, our reconstruction takes directly the object shifts into account in the imaging model, instead of modifying the data and registering the images as in~\cite{shroff_lateral_2010,chetty_structured_2012}. This enables us to minimize noise amplification using the Bayesian framework~\cite{orieux_fast_2017}. Secondly,  unlike the 2D SIM approaches, which achieve both OS and SR by using \emph{ad hoc} processing such as frequency component weighting~\cite{wicker_phase_2013,oholleran_optimized_2014,shaw_high_2015,muller_open-source_2016} or spectral merging~\cite{lukes_threedimensional_2014,hoffman_superresolution_2017}, our proposed method intrinsically enables both OS and SR. This is performed by explicitly considering the 3D nature of the object in the imaging model, even though the data are 2D, as in~\cite{jost_optical_2015}. In the latter paper, even though a whole 3D volume is reconstructed, only the in-focus slice of the volume is accurately recovered and the retrieved out-of-focus information is discarded. 
In contrast, our approach reconstructs a 2-layer object composed of the in-focus layer that contains the super-resolved information and a defocused layer into which the out-of-focus contribution is rejected, thus significantly reducing the computing cost.
Furthermore, our method is able to automatically adjust the reconstruction parameters for regularization. 

In Section~\ref{sect_method}, we present the 2-layer SIM imaging model that we exploit in an unsupervised Bayesian framework to obtain object reconstructions with OS and SR jointly. In Section~\ref{sect_simu}, we quantitatively assess the performance of our method in both OS and SR and compare it with fairSIM~\cite{muller_open-source_2016}, a state-of-the-art 2D SR-SIM method. Finally, we validate our approach on open-access experimental microscopy data~\cite{fairsim_parameter} in Section~\ref{sect_experimental}.


\section{Method}\label{sect_method}

We propose a general SIM reconstruction method suitable for moving object that is able to jointly perform super-resolution (SR) and optical sectioning (OS) in an unsupervised Bayesian framework. As such, the proposed method is tailored for \emph{in vivo} retinal imaging, \notevf{where eye movements induce inter-frame shifts of the retina}. It is based on a sound imaging model that aims to simulate the acquired images from the moving 3D object that is observed. This imaging model is then inverted using a Maximum a Posteriori (MAP) approach to reconstruct the super-resolved and optically sectioned slice of the object.

\subsection{SIM imaging model}\label{sect_model}

Let us firstly introduce some notations used to describe the imaging process.
We denote by $o$ the biological sample that is incoherently observed through an optical system. It can refer to the 3D fluorophore density in fluorescence microscopy or to the 3D reflectance map in retinal imaging for instance, and will be called object in the remainder of the paper.
$(x,y,z)$ are spatial coordinates in object space where $(x,y)$ are the transverse coordinates and $z$ is the axial coordinate, which quantifies the defocus from the object focal plane. In this paper, the defocus will be expressed using the axial normalized coordinate $u$ defined as in~\cite{wilson_resolution_2011,gruppetta_theoretical_2011}, by:
\begin{equation}\label{eq_def_u}
    u=8.(\pi/\lambda).z.n\sin^2(\alpha /2)
\end{equation}
where $n.\sin(\alpha)$ is the numerical aperture and $\lambda$ is the imaging wavelength.
It can also be related to the amount of defocus aberration $a_4$ expressed in radian rms in the Zernike polynomial decomposition of optical aberrations \cite{Noll_zernike_76} by $u=4\sqrt{3}.a_4$.
We assume that the object $o$ is moving in an uncontrolled way and that the object motion only implies inter-frame shifts in the transverse plane $(Oxy)$. This is a fair assumption in imaging modalities 
such as adaptive optics flood-illumination retinal imaging.

%
A set of illumination patterns $\{m_j\}_j$ is sequentially projected onto the observed sample~$o$.
Let us consider the j-th acquired image $\mathbf{i}_j$ by a camera at the image focal plane. Due to its motion, the object is shifted in the j-th image by $\delta_j(x,y)=\delta(x-x_j,y-y_j)$ where $\delta$ is the 2D Dirac delta function.
The imaging process can be represented by a 3D convolution of the observed sample with the incoherent Point-Spread-Function (PSF) of the optical system. Here, because we only record images in one plane, we prefer to describe it with 2D convolutions of PSF and object layers integrated along the optical axis. 
The image $\mathbf{i}_j$ is then the sum of 2D convolutions of each object layer $o_z(x,y)=o(x,y,z)$ shifted by $\delta_j(x,y)$
, modulated by the j-th illumination pattern $m_{j,z}(x,y)=m_j(x,y,z)$ with the PSF layer $h_z$ conjugated to depth $z$, and integrated along the optical axis :
\begin{equation}\label{eq_model_continu}
    \mathbf{i}_j(k,l)=\big[ \int_z h_z \ast (m_{j,z}\times (o_z \ast \delta_j)) \ddroit{z} \big]_{\Sh}(k,l) + \mathbf{n}_j(k,l)
\end{equation}
where $(k,l)$ refers to the camera pixel indices, $\ast$ and $\big[.\big]_{\Sh}$ depicts the 2D convolution product and the sampling operator at camera pixels respectively. The detector noise $\mathbf{n}_j$ is a mixture of readout noise which is homogeneous white Gaussian and of photon noise which follows Poisson statistics. We assume that the number of collected photons is high enough to reasonably approximate this mixture by a spatially independent, inhomogeneous centered Gaussian noise~\cite{mugnier_mistral_2004} with variance map $\boldsymbol{\sigma}_j^2$.
Eq.~(\ref{eq_model_continu}) needs to be discretized to make the imaging model computable. The integral is approximated by a Riemann sum involving the discretized versions of $h_z$, $m_{j,z}$ and $o_z$, denoted in bold, so that our discrete imaging model is:
\begin{equation}\label{eq_model_discret}
    \mathbf{i}_j(k,l) = \sum_z \big[\mathbf{h}_z \star (\mathbf{m}_{j,z}. t_j[\mathbf{o}_z]) \big]_{\boldsymbol{\mathrm{III}}}(k,l) + \mathbf{n}_j(k,l)
\end{equation}
where $\star$ depicts the discrete 2D convolution product and $t_j[.]$ is a subpixel shift operator that computes the shifted discretized object.
\notevf{The downsampling operator $[.]_{\boldsymbol{\mathrm{III}}}$ was introduced in order to make it explicit that the object and the PSFs may be defined on a grid that is finer than the sampling grid of the image. For instance, if the images are just Shannon-Nyquist sampled, then $\mathbf{h}_z$, $\mathbf{m}_{j,z}$ and $\mathbf{o}_z$ must be defined on a grid twice as fine as that of $\mathbf{i}_j$ so that the frequency support of the reconstructed super-resolved object is large enough.
On the contrary, if the images are oversampled by a factor 2 w.r.t. the optical cutoff frequency, $\mathbf{o}_z$ and the $\{\mathbf{i}_j\}_j$ can be sampled on the same grid.}

We can rewrite our forward model from Eq.~(\ref{eq_model_discret}) in a classical matrix framework for computing the imaging model as in \cite{orieux_bayesian_2012}. 
The object layers $\mathbf{o}_z$, are written as a vector where the lines of the image are column stacked.
The shifting of the object by $\delta_j$ is expressed as the multiplication of the object layer $\mathbf{o}_z$ with a matrix $\mathbf{T}_j$. The modulation by the illumination pattern is modelled by the product between the shifted object and the diagonal matrix $\mathbf{M}_{j,z}$ whose diagonal elements correspond to the illumination pattern $\mathbf{m}_{j,z}$. The convolution product is then written as a multiplication of the modulated and shifted object with the matrix $\mathbf{H}_{z}$ which is a Toeplitz-block-Toeplitz matrix. Its first line corresponds to the 2D PSF $\mathbf{h}_z$. The downsampling operator is expressed as a matrix $\mathbf{D}$.
The full expression of the forward model is obtained by successively applying each of these matrices to $\mathbf{o}_z$, thus Eq.~(\ref{eq_model_discret}) can be rewritten as :

\begin{equation}\label{eq_model_bayes}
    \mathbf{i}_j = \sum_z \mathbf{D}\mathbf{H}_z \mathbf{M}_{j,z} \mathbf{T}_j\mathbf{o}_z + \mathbf{n}_j
\end{equation}

Considering the whole set of data $\{\mathbf{i}_j\}_{1\leq j \leq N}$ with $N$ the number of images, we have a system of $N$ equations given by Eq.~(\ref{eq_model_bayes}), whose unknowns are the object layers $\{\mathbf{o}_z\}_z$ and the object shifts $\{\mathbf{T_j}\}_j$.
The main issue to discuss is then: how many object layers should be reconstructed in order \notevf{to jointly} achieve OS and SR for the focused layer $\mathbf{o_0}$? To answer to this question, it is important to understand the physical origin of SIM's OS.



\begin{figure}[htbp]
\centering\includegraphics[width=0.8\linewidth]{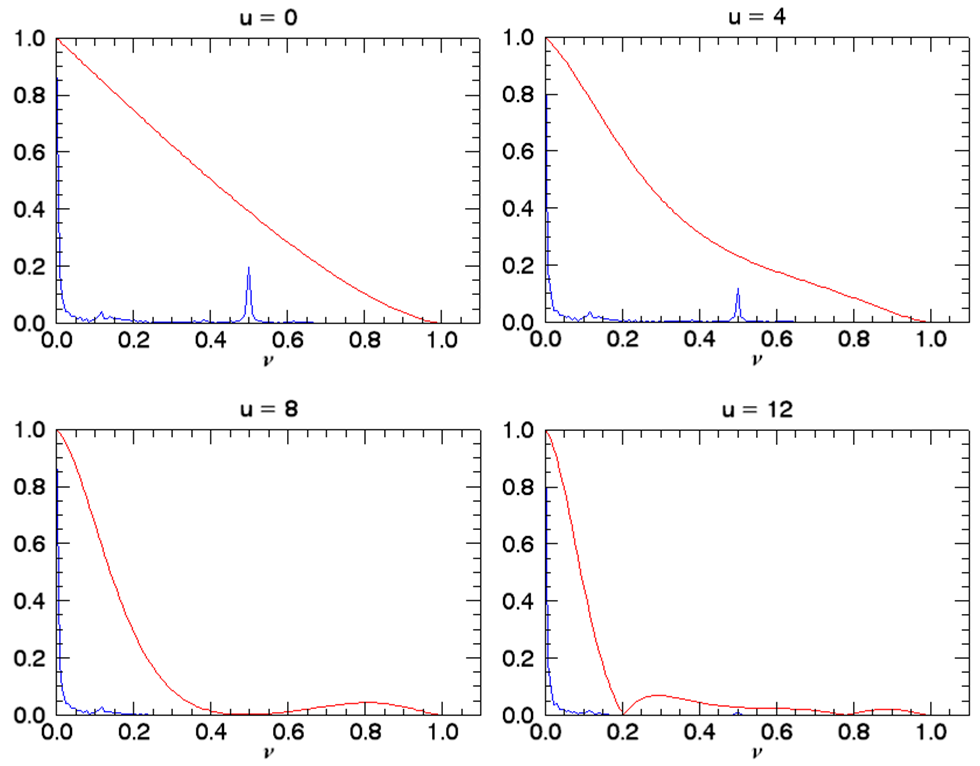}
\caption{\small{
Contribution of a sinusoidally illuminated object layer in Fourier space to the image plane (in blue) through an unaberrated incoherent optical system whose modulation transfer function (MTF) $\abs{\Tilde{\mathbf{h}}_u}(\nu)$ is superimposed in red for different defocus values $u$. The abscissa coordinate, $\nu=f/f_c$, is the radial normalized spatial frequency. The modulated spectra are normalized by their value at zero frequency.
The sinusoidal illumination produces a modulated replica of the object layer's spectrum shifted around the modulation frequency $\nu_m=0{.}5$ on the plots. The amplitude of its contribution to the image depends on the value of the MTF at the modulation frequency. 
}}
\label{fig_mtf}
\end{figure}

The optical sectioning capability of SIM, evaluated in~\cite{wilson_resolution_2011}, is based on the fact that the contrast of the fringes on each object layer decreases in the acquired images with its distance \notevf{from} the object focal plane. 
In the Fourier domain, the modulation of each object layer by the illumination pattern translates into the convolution of the object layer's spectrum with the illumination pattern spectrum. In the case of a sinusoidal modulation, the spectrum of each modulated object layer is thus the object layer's spectrum with two added replicas of itself shifted around the carrier frequency $\pm f_m$ of the modulation. During the image formation process, each modulated layer's spectrum is attenuated by the OTF $\Tilde{\mathbf{h}}_z$ in Fourier space, as \notevf{exemplified} in Fig.~\ref{fig_mtf}, and the acquired image spectrum is the sum of each of these contributions.
Of course, the higher the defocus $u$ of a given object layer is, the more the medium and high spatial frequencies are attenuated in the images by the modulation transfer function (MTF) (Fig.~\ref{fig_mtf}, red plots), and consequently the less contrasted its spectrum's replicas are (Fig.~\ref{fig_mtf}, blue plots). Thus, the modulated spectrum components in the acquired SIM images essentially come from the object layers near the focal plane and this enables OS.
In particular, if we choose a modulation frequency equal to
$\nu_m(=f_m/f_c)=0{.}5$, which has been found to be the optimal choice for OS~\cite{wilson_resolution_2011}, the modulation completely vanishes for a defocus distance $z_d$ corresponding to $u=8$ (or $a_4=2/\sqrt{3}$ radian rms defocus) as shown in Fig.~\ref{fig_mtf}. And in practice, the modulation is also negligible for planes more defocused than $z_d$, because the corresponding OTF is even more low-pass.

Considering the above, we choose in this paper to model the object as 2 planes only, the focused plane $z=0$ and a defocused plane $z=z_d$ and to reconstruct a 2-plane object $\mathbf{o}=(\mathbf{o}_0,\mathbf{o}_d)$. This is obviously the minimum number of planes in order to perform OS, i.e, reject \notevf{defocused} light into $\mathbf{o_d}$, and this allows for a fast algorithm as it keeps the number of unknowns to a minimum.
We have checked, as shown in Section~\ref{sect_simu}, that if we choose the defocused plane to be $z=z_d$, in practice object contributions that are more defocused than $z=z_d$ are incorporated into the estimated $\hat{\mathbf{o}}_d$ at plane $z_d$ so it is unnecessary to add more defocused planes to be estimated.
We have also checked that choosing \notevf{a distance $z < z_d$ as the defocused layer} to be reconstructed is suboptimal, as it results in a poorer sectioning than $z=z_d$ due to a poorer disambiguation of each layer's contribution. Thus, in this paper, we shall reconstruct a 2-plane object lying at $z=0$ and $z=z_d$.
We note $\mathbf{M}_{j,d}$, and $\mathbf{H}_{d}$ respectively the modulation and convolution matrices corresponding to the defocused layer $\mathbf{o}_d$.
Similarly, we note $\mathbf{M}_{j,0}$ and $\mathbf{H}_{0}$ respectively, the modulation and convolution matrices for the in-focus layer $\mathbf{o}_0$. This leads us to the following expression \notevf{for} the 2-layer imaging model:
\begin{align}
    &\mathcal{M}_j(\mathbf{o}_0,\mathbf{o}_{d})
    = \mathbf{D}\mathbf{H}_0 \mathbf{M}_{j,0} \mathbf{T}_j\mathbf{o}_0 + \mathbf{D}\mathbf{H}_{d} \mathbf{M}_{j,d} \mathbf{T}_j\mathbf{o}_{d} + \mathbf{n}_j \label{eq_model_reduit}
\end{align}

The novelty of this model is that the out-of-focus contribution and the object shifts are explicitly taken into account through the defocused layer $\mathbf{o}_{d}$ and the translation matrices $\mathbf{T}_{j}$. By inverting the model within the Bayesian framework as described in the next section, we will be able to reconstruct the object focused slice with SR and OS jointly in the in-focus layer $\mathbf{o}_{0}$, while properly rejecting the out-of-focus contribution into the defocused layer $\mathbf{o}_{d}$, even if the object moves between frames.
Thus, our reconstruction method is nicknamed BOSSA-SIM, which stands for Bayesian Optical Sectioning and Super-resolution Algorithm for SIM.

\subsection{Reconstruction method}\label{section_reconstruction}

Before inverting the forward model given by Eq.~(\ref{eq_model_reduit}), the first step consists in determining the model parameters. \notevf{PSF} and illumination pattern calibrations, which are discussed in Subsection~\ref{sect_calib_instru}, enable one to set the modulation matrices $(\mathbf{H}_0,\mathbf{H}_{d})$ and convolution matrices $\{(\mathbf{M}_{j,0},\mathbf{M}_{j,d})\}_j$. 
As for the object shifts $\{\delta_j\}_j$ from which we define the shifting matrices $\{\mathbf{T}_j\}_j$, they are set to 0 if the observed sample is static as \notevf{is} generally the case in microscopy. Otherwise, they are estimated from the acquired SIM images using a subpixel shift registration algorithm described in Subsection~\ref{sect_shift}.

Once the model parameters \notevf{are} determined, the 2-layer object $\mathbf{o}=(\mathbf{o}_0,\mathbf{o}_{d})$ is estimated within the Bayesian MAP framework. 
The topic of Bayesian approaches for image estimation is well documented \cite{Idier08a}. 
The MAP estimator of the 2-layer object $\mathbf{o}$ is given by maximizing the posterior likelihood of $\mathbf{o}$, knowing the acquired SIM images $\{\mathbf{i}_j\}_{1 \leq j \leq N}$:
\begin{equation}\label{eq_map_proba}
    \hat{\mathbf{o}}=\arg \max_\mathbf{o} p(\mathbf{o}|\mathbf{i}_1, \mathbf{i}_2, ..., \mathbf{i}_N)
\end{equation}
where the symbol $\hat{.}$ denotes the estimated value.

Maximizing the posterior likelihood is equivalent to minimizing the \textit{neg-log-likelihood} $J(\mathbf{o})=-\ln p(\mathbf{o}|\mathbf{i}_1, \mathbf{i}_2, ..., \mathbf{i}_N)$. As the observed images $\mathbf{i}_j$ are independently corrupted by a white centered Gaussian noise \notevf{with} variance map $\boldsymbol{\sigma}_j^2$, one can show using Bayes'rule that the criterion $J$ is the sum of a weighted least square metric and a regularization term $R(\mathbf{o})=-\ln p(\mathbf{o})$ that embodies the object prior information:
\begin{equation}\label{eq_map_criterion1}
    J(\mathbf{o}) = \frac{1}{2} \sum_{j=1}^N \norme {\frac{\mathbf{i}_j - \mathcal{M }_j(\mathbf{o})}{\boldsymbol{\sigma}_j}}_2^2 + R(\mathbf{o})
    = \frac{1}{2} \sum_{j=1}^N \sum_{k,l} \abs{ \frac{\mathbf{i}_j(k,l) - \mathcal{M}_j(\mathbf{o})(k,l)}{\boldsymbol{\sigma}_j(k,l)}}^2 + R(\mathbf{o})
\end{equation}
where $\norme{.}_2^2$ is the square of the Euclidian norm of its argument and $\mathcal{M}(\mathbf{o})$ is the forward model of Eq.~(\ref{eq_model_reduit}).

The regularization term $R(\mathbf{o})$, which prevents uncontrolled noise amplification, uses the following prior: we assume that the 2 layers of the object $(\mathbf{o}_0,\mathbf{o}_{d})$ are independent and that they each follow centered Gaussian statistics.
We thus can write $R(\mathbf{o})=R_{\mathbf{o}_0}(\mathbf{o}_0)+R_{\mathbf{o}_{d}}(\mathbf{o}_{d})$ where $R_{\mathbf{o}_0}(\mathbf{o}_0)$ and $R_{\mathbf{o}_{d}}(\mathbf{o}_{d})$ are the respectively in-focus and out-of-focus object priors. They can be written in Fourier space as~\cite{Conan-a-98}:
\begin{equation}\label{eq_regu_gauss}
    R_{\mathbf{o}_q}(\mathbf{o}_q)=\frac{1}{2} \sum_{f}{\frac{\abs{\Tilde{\mathbf{o}}_q(f)}^2}{ \mathbf{S}_{\mathbf{o}_q}(f)}}
\end{equation}
where the index $q$ refers either to "0" or "d", the tilde denotes the 2D discrete Fourier transform and $f$ is the \notevf{2D} spatial frequency. $\mathbf{S}_{\mathbf{o}_q}$ is the power spectral density (PSD) of the layer $\mathbf{o}_q$.

By injecting Eq.~(\ref{eq_regu_gauss}) into Eq.~(\ref{eq_map_criterion1}), we obtain the expression of the MAP criterion to be minimized under positivity constraint:
\begin{equation}\label{eq_map_criterion}
    J(\mathbf{o})=\frac{1}{2} \sum_{j=1}^N \norme {\frac{\mathbf{i}_j - \mathcal{M}_j(\mathbf{o})}{\boldsymbol{\sigma}_j}}_2^2 +     \frac{\lambda}{2} \left [ \sum_{f}{\frac{|\Tilde{\mathbf{o}}_0(f)|^2}{ \mathbf{S}_{\mathbf{o}_0}(f)}} + \sum_{f}{\frac{|\Tilde{\mathbf{o}}_{d}(f)|^2}{ \mathbf{S}_{\mathbf{o}_{d}}(f)}} \right ]
\end{equation}
where $\lambda$ is a regularization parameter which should be set to $1$ for a nominal regularization.
The noise variance maps $\boldsymbol{\sigma}_j^2$ and the PSDs of each object layer, denoted $\mathbf{S}_{\mathbf{o}_0}$ and $\mathbf{S}_{\mathbf{o}_{d}}$, are estimated from the data in an unsupervised way before minimizing the criterion $J$. The details of this estimation step are reported in Subsection~\ref{sect_hyperparam_estimation}.
As there is no analytical expression of the 2-layer object $\mathbf{o}$ that minimizes the criterion $J$, the minimization is numerically computed using the VMLM-B method~\cite{Thiebaut_b_2002}.

%
%



\subsection{Implementation}\label{sect_implementation}

\subsubsection{Instrument parameter calibration}\label{sect_calib_instru}

The calibration of the PSFs and especially the illumination patterns needs to be carefully done in order to avoid reconstruction artefacts~\cite{schaefer_l._h._structured_2004}. Depending on the optical system and the object medium, we can either consider an ideal diffraction-limited or an aberrated in-focus PSF $\mathbf{h}_0$. In the latter case, the PSF can be derived from the optical aberrations if they are measured~\cite{Noll_zernike_76}, or from \notevf{empirical} approaches if not. We use, as in the fairSIM quickstart guide available in~\cite{fairsim_parameter}, an exponential attenuation of the ideal OTF to account for optical aberrations in the experimental microscopy data that we process in Section~\ref{sect_experimental}.
As for the out-of-focus PSF $\mathbf{h}_{d}$, we consider a defocused version of PSF $\mathbf{h}_0$ with a defocus of $u=8$ as explained in Subsection~\ref{sect_model}.

\subsubsection{Subpixel shift estimation}\label{sect_shift}

In the case of a static object, the object shifts are simply set to 0. If the object is moving, the object shifts are estimated from the acquired SIM images. The estimation method assumes that the motion implies only shifts as is customary in flood-illumination retinal imaging, and is an adaptation of the image registration algorithm presented in~\cite{blanco_registration_2014}, which allows \notevf{to register all images jointly rather than pairwise, within the MAP framework}. The reader is referred to the original paper for more implementation details. The original algorithm was tailored for conventional retinal imaging (without structured illumination). 
In order to apply it \notevf{to} SIM imaging, we filter each SIM image to cut off the spatial frequencies around the modulation frequencies of the illumination pattern in Fourier space. In practice, the filtering is performed by multiplying \notevf{the SIM image in Fourier space} with a binary mask whose values are set to 0 on a 3 pixel radius around the modulation frequencies and set to 1 elsewhere. This pre-processing step allows one to approximate the filtered images as conventional images, on which we next apply the image registration algorithm from~\cite{blanco_registration_2014}. This approximation leads to additional errors in the estimation of the object shifts but we show in simulation (Section~\ref{sect_simu}) that they are small enough not to degrade the reconstruction performance.

\subsubsection{Hyper-parameter estimation for unsupervised reconstruction}\label{sect_hyperparam_estimation}

In order to compute the MAP criterion from Eq.~(\ref{eq_map_criterion}), the noise variance maps, the PSDs of each object layer and the regularization parameter $\lambda$ need to be determined. We propose a method to automatically estimate them.

The noise variance map $\boldsymbol{\sigma}_j^2$ at every pixel $(k,l)$ is assumed to be the mixture of a photonic noise approximated by a white Gaussian noise of variance $\boldsymbol{\sigma}_{j,ph}^2(k,l)$ and a readout noise which is white Gaussian of variance $\boldsymbol{\sigma}_{j,det}^2(k,l)$. The readout noise is characterized through a calibration of the detector whereas the variance of the photon noise is estimated using $\hat{\boldsymbol{\sigma}}_{j,ph}^2(k,l)= \max (\mathbf{i}_j(k,l),0)$ assuming that the camera gain has been properly calibrated. The noise variance maps are then estimated using $\hat{\boldsymbol{\sigma}_j}^2(k,l)=\hat{\boldsymbol{\sigma}}_{j,ph}^2(k,l) + \hat{\boldsymbol{\sigma}}_{j,det}^2(k,l)$.
If the camera isn't calibrated, then the noise variance maps are approximated by a mean variance which can be estimated using an unsupervised method presented below.  

The PSD of each layer, $\mathbf{S}_{\mathbf{o}_0}$ and $\mathbf{S}_{\mathbf{o}_d}$ as well as the noise mean variance $\sigma^2$ are estimated from a widefield image of the object. For a 2-layer object, each layer contributes to the image PSD whose model is given by :
\begin{equation}\label{eq_model_dsp} 
    \mathbf{S}_\mathbf{i} 
    = \abs{\Tilde{\mathbf{h}}_0}^2 \mathbf{S}_{\mathbf{o}_0} + \abs{ \Tilde{\mathbf{h}}_d}^2 \mathbf{S}_{\mathbf{o}_d} + S_{\mathbf{n}} \\
\end{equation}
where $S_{\mathbf{n}}$ is the noise PSD which is, up to a multiplicative constant, equal to the noise variance.
In order to estimate the PSD of the two layers $\mathbf{o}_0$ and $\mathbf{o}_{d}$, we assume that their PSDs are the same apart from a scaling factor that depends on the amount of photons backscattered by each layer. If we note $\alpha$ the portion of photons collected by the detector that is attributed to the in-focus layer, then a ratio $1-\alpha$ of the collected photons comes from the out-of-focus layer.
Then we have :
\begin{equation}\label{eq_approx_psd}
    {\mathbf{S}}_{\mathbf{o}_0}= \alpha^2 {\mathbf{S}}_\mathbf{o} \qquad \text{and} \qquad {\mathbf{S}}_{\mathbf{o}_{d}}= (1-\alpha)^2 {\mathbf{S}}_\mathbf{o}
\end{equation}
where $\mathbf{S}_\mathbf{o}$ can be called the single-layer object PSD. Eq.~(\ref{eq_approx_psd}) can be injected in Eq.~(\ref{eq_model_dsp}) to obtain :
\begin{equation}\label{eq_model_dsp_2} 
    \mathbf{S}_\mathbf{i} 
    = \left [ \alpha^2 \abs{\Tilde{\mathbf{h}}_0}^2 + (1 - \alpha)^2 \abs{\Tilde{\mathbf{h}}_d}^2 \right ] \mathbf{S}_{\mathbf{o}} + S_{\mathbf{n}} \\
\end{equation}

From Eq.~(\ref{eq_model_dsp_2}), we can deduce that the image PSD is the same as the one obtained for a 2D object observed with an imaging system of equivalent MTF $\abs{\Tilde{\mathbf{h}}_{eq}}= \sqrt{ \alpha^2 \abs{\Tilde{\mathbf{h}}_0}^2 + (1 - \alpha)^2 \abs{\Tilde{\mathbf{h}}_d}^2}$. By using the unsupervised method from~\cite{Blanc-a-03b} designed for a 2D object, giving the MTF $\abs{\Tilde{\mathbf{h}}_{eq}}$ in input, we obtain from the image an estimated 3-parameter PSD $\hat{\mathbf{S}}_\mathbf{o}$ of the object and the noise mean PSD $S_\mathbf{n}$~\cite{Conan-a-98}. We then easily derive the estimation of each layer's PSD from Eq.~(\ref{eq_approx_psd}) given $\alpha$.
The parameter $\alpha$ should depend on the observed sample and the imaging setup. In practice we choose $\alpha = 0.5$, assuming that the in-focus layer and the defocused layer scatter the same amount of photons toward the detector. We show in simulations and on experimental microscopy data that this value yields good results.

The regularization parameter $\lambda$ for the minimization of the MAP criterion of Eq.~(\ref{eq_map_criterion}) should be fixed to $1$ for nominal regularization. In the reconstructions that are performed in the next sections, we use $\lambda=0.3$, i.e. a slight under-regularization in order to take into account the fact that the positivity constraint brings some additional regularization.



\subsubsection{Object reconstruction: MAP criterion minimization}

\notevf{Once the instrument parameters, the object shifts and the noise and object PSDs have been determined as described above, the 2-layer object $\mathbf{o}$ is estimated by minimizing the MAP criterion $J(\mathbf{o})$ from Eq.~(\ref{eq_map_criterion}).
To perform this minimization we use a fast iterative gradient-based method called VMLM-B~\cite{Thiebaut_b_2002}. It requires an initial guess of the unknowns $(\mathbf{o}_0,\mathbf{o}_{d})$ and it computes repeatedly the MAP criterion $J$ and its gradient until convergence. The minimization is automatically stopped when the evolution of the estimated object from one iteration to the next is close to machine precision. }

\notevf{As the imaging model of Eq.~(\ref{eq_model_reduit}) is a linear function of all object pixel values, then the data fidelity term is quadratic and thus a convex function of the object pixel values. The regularization metrics are also quadratic functions of the object pixel values, and thus convex. Additionally, the chosen regularization metric of Eq.~(\ref{eq_regu_gauss}) is a weighted quadratic sum of all object frequency components (with only non-zero weights) and is thus strictly convex. We then conclude that criterion $J$ is strictly convex as a sum of a convex data fidelty term and a strictly convex regularization metric. Hence, it does not have any local minimum~\cite{nocedal2006numerical}, and we can safely initialize the unknowns to any value, e.g. zero.}

\begin{figure}[htb]
    \centering
    \includegraphics[width=1.\linewidth]{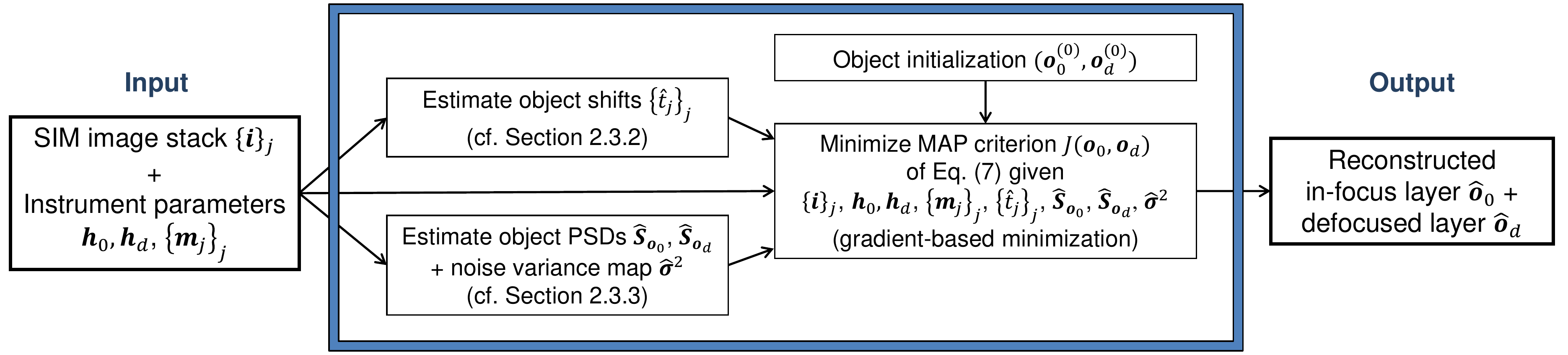}
    \caption{Block diagram of the BOSSA-SIM algorithm}
    \label{fig_algo}
\end{figure}

The whole algorithm is summarized in the block diagram of Fig.~\ref{fig_algo}.





\section{Performance assessment by simulations}\label{sect_simu}

\subsection{Simulation conditions}\label{subsect_simucond}

In order to validate this approach, we have simulated a 16-layer object. Each layer is composed of a radial resolution target of the form $f(\rho,\theta)=1+\cos{(40\theta)}$ where $(\rho,\theta)$ are the polar coordinates. This kind of object allows one to visually evaluate the SR and OS simultaneously. The object layers are equally spaced along the optical axis so as to have the first one at the object focal plane and the other ones gradually more defocused with a defocus step of $u=1$ in normalized axial coordinates.
This object layout was chosen because it allows us to conveniently assess the performance of the reconstructions in both OS and SR, as we will see later.

We consider that a physical single spatial frequency grid structure at spatial frequency $\vec{f_m}$ is imaged onto the object to produce the structured illumination and the object is imaged on a camera through the same optical system as the illumination one. 
In this case, the contrast of the projected modulation pattern in object plane at defocus z is given by the value $C_z(f_m)$ of the MTF of the optical system at the spatial frequency of the modulation $f_m=||\vec{f_m}||$ and defocus $z$.
The simulated illumination pattern projected onto the object layer at defocus $z$ is then:
\begin{equation}\label{eq_simu_illu}
    \mathbf{m}_z(\vec{r})=1+C_z(f_m) \cos(2\pi \vec{f}_m.\vec{r}+\phi)
\end{equation}
where $\vec{r}=(k,l)$ refers to the pixel indices, and $\phi$ specifies the relative position of the illumination pattern to the object frame.
Horizontal and vertical sinusoidal illumination patterns with a modulation frequency $f_m$ equal to half the optical cutoff frequency $f_c$ are used, leading to a theoretical SIM cutoff frequency of $f_m+f_c=1.5 f_c$. Although higher modulation frequency can be used to achieve better resolution, this value was chosen because it exhibits optimal OS~\cite{chetty_structured_2012}. 


The PSF of the optical system is assumed to be diffraction-limited. Thus we model the in-focus PSF $\mathbf{h}_0$ of the optical system by an ideal diffraction-limited Airy disk~\cite{goodman1996introduction}.
The defocused PSFs $\mathbf{h}_{z,z \neq 0}$ are then derived from $\mathbf{h}_0$ by adding the corresponding amount of defocus aberration. To fulfill the Shannon-Nyquist sampling theorem during object sampling and reconstruction, the diffraction-limited PSF is oversampled by a factor of 3 (\emph{i.e.}, the Nyquist frequency is three times higher than the optical cutoff frequency). This allows to conveniently use the same spatial sampling for the simulated object, the simulated images and the SIM reconstructions.

\begin{figure}[htb]
\centering\includegraphics[width=0.8\linewidth]{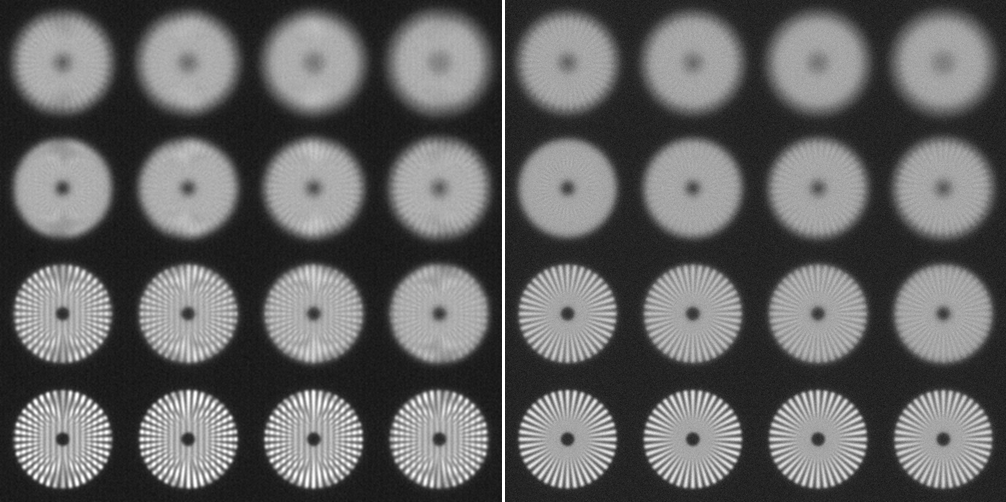}
\caption{\small Simulated images for a SNR of 10: SIM images with vertical modulation pattern (left) and conventional widefield image (right). The in-focus radial target is located in the bottom-left corner, then the other targets are progressively defocused from left to right and from bottom to top with a defocus step of $u=1$.}
\label{fig_simu_data}
\end{figure}

The widefield image and the SIM images $\{\mathbf{i}_l\}_l$ of size 1024x1024 pixels are then simulated using Eq.~(\ref{eq_model_discret}) for different SNRs (Signal-to-Noise ratios). We define the SNR as the mean of the image over the standard deviation of the noise and we assume that the detector noise is an additive white gaussian noise. 
Examples of simulated images for a SNR of 10 are displayed in Fig.~\ref{fig_simu_data} and show how the different radial targets become more and more blurred as the defocus grows.

Two sets of simulations were conducted. In the first one, the object is static and illuminated with 3 different phases ($\phi=0$, $\phi=2\pi/3$ and $\phi=4\pi/3$) for each of the 2 directions (horizontal and vertical) of the pattern, making a total of $3\times2=6$ SIM images.

\notevf{The second case aims at simulating a moving object such as the retina. Eye movements are well studied and quite complex~\cite{martinez-conde_role_2004} and induce inter-frame shifts of the retina during the image acquisition. For SIM imaging, the most important requirement on the retinal shifts is that they span more than one spatial period of the modulation pattern so that the corresponding phase between the object and the pattern covers $[0,2\pi]$~\cite{shroff_lateral_2010}, a condition that is always verified in practice. In our simulation, the object is randomly shifted in the transverse plane following a uniform law of bounds $[-32,32]$ pixels, which is notably larger than the period of the chosen modulation pattern, $2/f_c$ or $12$ pixels.}
The phase of the illumination patterns is fixed ($\phi=0$) between images, and it is the object movements that introduce the required phase shift diversity for reconstruction. $7$ images of the object are simulated for each of the 2 pattern directions to ensure a suitable diversity of the random phases as shown in~\cite{chetty_structured_2012}. The total number of considered SIM images is then $14$.



\subsection{Reconstruction methods}

We perform two types of reconstruction: the proposed BOSSA-SIM reconstructions presented in Section~\ref{sect_method} and the open-source fairSIM method~\cite{muller_open-source_2016} that enables 2D SIM reconstructions with SR and OS.
The BOSSA-SIM reconstructions were performed using the illumination patterns defined in Eq.~(\ref{eq_simu_illu}) and a diffraction-limited in-focus PSF $\mathbf{h}_0$. The reconstructions were then carried out as explained in Subsection~\ref{sect_implementation}.

The fairSIM reconstruction consists of 4 steps of processing in Fourier space: the different spectral components are firstly extracted, taking into account the OTF attenuation, then shifted into place; they are next combined through a weighted generalized Wiener filter and the result is finally apodized with a selected OTF profile to avoid ringing artefacts. Such \notevf{a} method requires several reconstruction parameters to be adjusted empirically: the shape and strength of the OTF attenuation that suppresses frequency components with poor axial resolution to achieve OS, the Wiener parameter $w$ for regularization and the shape of the apodization function.
To perform fairSIM reconstruction on the simulated data, an ideal OTF was selected and the OTF attenuation was set with the default values proposed by fairSIM (strength of $0.99$ and FWHM of $1.2$).
The Wiener parameter $w$ was chosen so as to ensure the same noise amplification in fairSIM and BOSSA-SIM reconstruction, thus allowing fair comparisons between the two methods.
The noise amplification is evaluated by computing the normalized reconstruction noise standard deviation $\sigma_{rec,norm}=\sigma_{rec}/m$, where $m$ is the mean intensity of the reconstructed in-focus target and $\sigma_{rec}$ is the reconstruction noise standard deviation, which is estimated on a background area dominated by noise. This leads to choosing $w=0.01$ for a SNR of 100 and $w=0.26$ for a SNR of 10.


\subsection{Performance assessment criteria}\label{sect_perf_criteria}

We define two performance criteria to assess the achieved OS and SR in the reconstructions. Thanks to the above choice for the regularization parameter of fairSIM, we are able to compare the performance of both reconstruction methods for the same noise amplification.

The OS is evaluated by studying the mean intensity axial profile $I(u)$ of the reconstructions as a function of the defocus $u$. As the object is composed of 16 radial targets each of which being at a specific defocus $u$ from $u=0$ to $u=15$ by steps of $1$ as illustrated in Fig.~\ref{fig_simu_data}, we compute the intensity axial profile by measuring the mean intensity of each radial target. In practice, we consider 16 circular regions of interest (ROIs), each corresponding to the positions of a radial target at a given defocus $u$ in the reconstructions and we compute the mean intensity over each ROI to obtain the mean intensity axial profile of the reconstructions.
The strength of the axial sectioning is then measured by the half-width $u_{1/2}$ of the mean intensity axial profile, as in~\cite{wilson_resolution_2011}.


The SR is evaluated by measuring the contrast of the reconstructed in-focus radial target located in the bottom-left corner of the image, as a function of the spatial frequency. As in~\cite{thesis_liu2018}, we compute the following modulation contrast function (MCF):
\begin{equation}
    C(f)=2.\Tilde{\mathbf{o}}_{rec}(f)/\Tilde{\mathbf{o}}_{rec}(0)
\end{equation}
where $f$ is the spatial frequency, $\Tilde{\mathbf{o}}_{rec}$ is the 1D Fourier transform of the reconstructed in-focus radial target interpolated along a centered circle of radius $40/(2.\pi.f)$ expressed in meters.
By comparing the MCF of the ground truth object and the MCF of the reconstructed one, we are able to assess how well the spatial frequency components of the object are reconstructed and up to which frequency. For each reconstruction, we define the maximal reconstructed frequency as the spatial frequency for which the MCF falls down below $0{.}2$.



\subsection{Results}

\subsubsection{Static object}

\begin{figure*}[htbp]
    \centering
    \begin{subfigure}[b]{0.32\linewidth}
        \centering
        \includegraphics[width=\columnwidth]{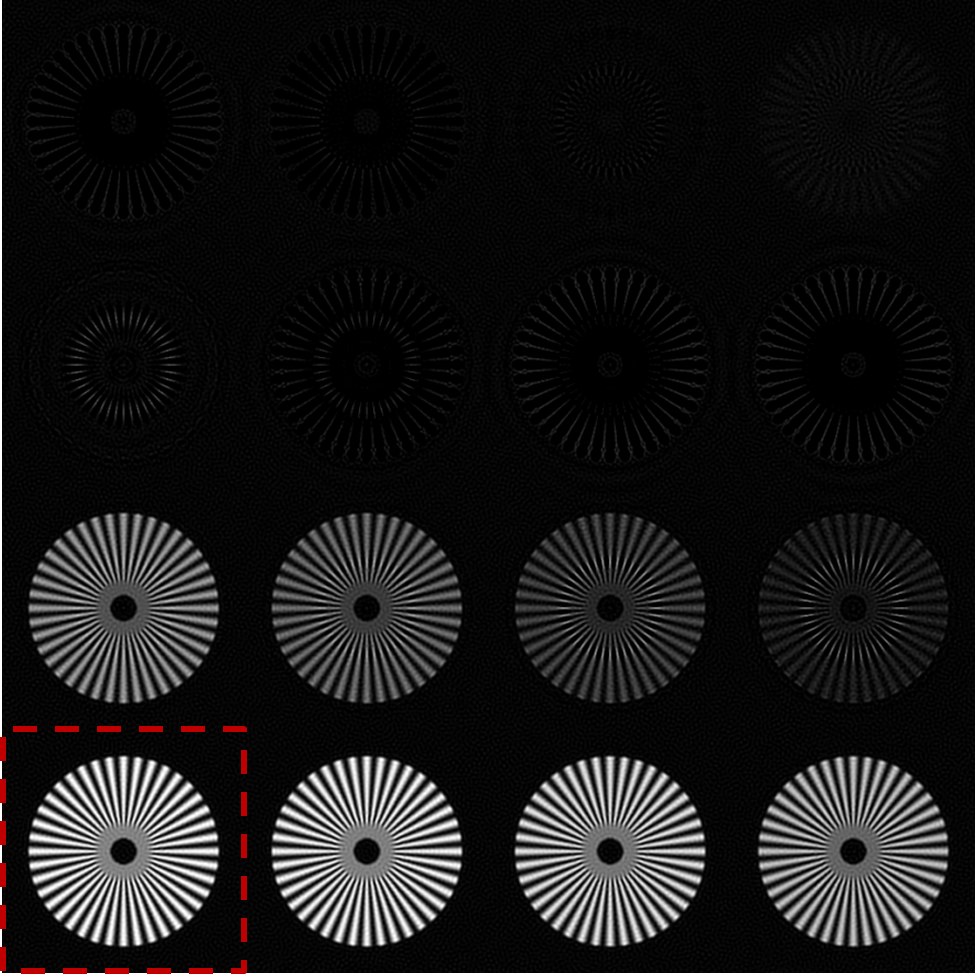}
        \caption{\centering In-focus layer $\hat{\mathbf{o}}_0$, SNR=100}
        \label{subfig:static_o0_100}
    \end{subfigure}
    ~
    \begin{subfigure}[b]{0.32\linewidth}
        \centering
        \includegraphics[width=\columnwidth]{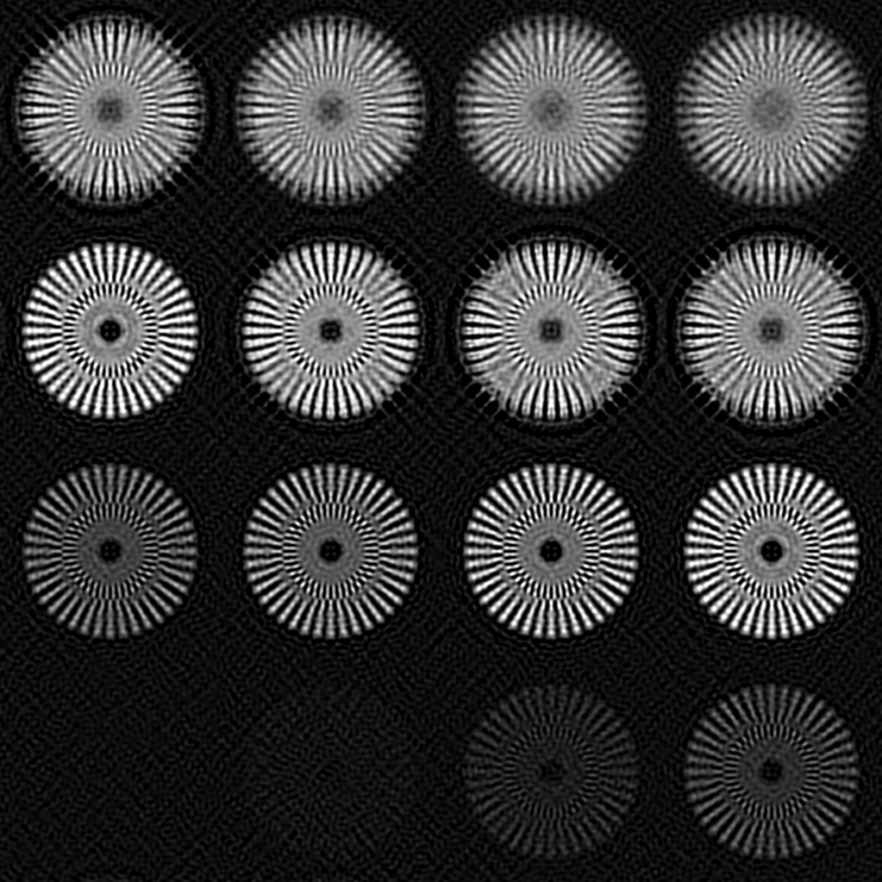}
        \caption{\centering Defocused layer $\hat{\mathbf{o}}_d$, SNR=100}
        \label{subfig:static_odefoc_100}
    \end{subfigure}
    ~
    \begin{subfigure}[b]{0.32\linewidth}
        \centering
        \includegraphics[width=\columnwidth]{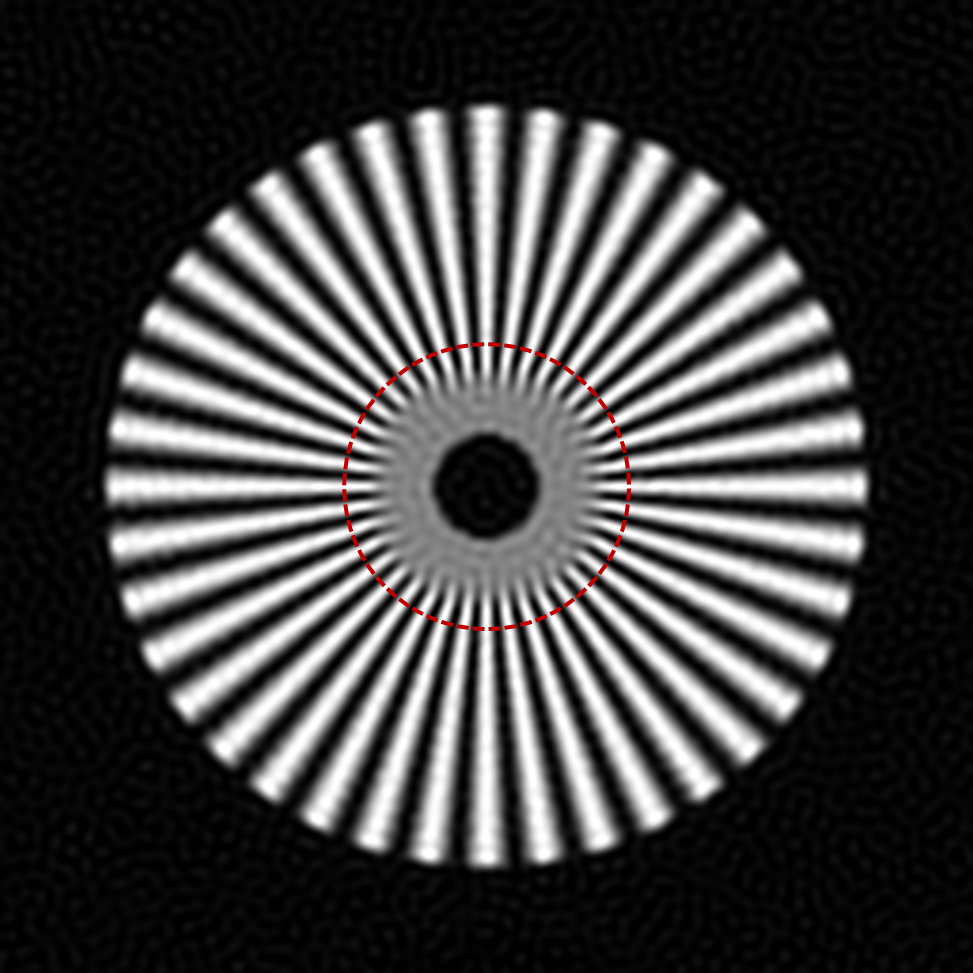}
        \caption{\centering Zoomed-in region, SNR=100}
    \end{subfigure}\vspace*{.5\baselineskip}
    
    \centering
    \begin{subfigure}[b]{0.32\linewidth}
        \centering
        \includegraphics[width=\columnwidth]{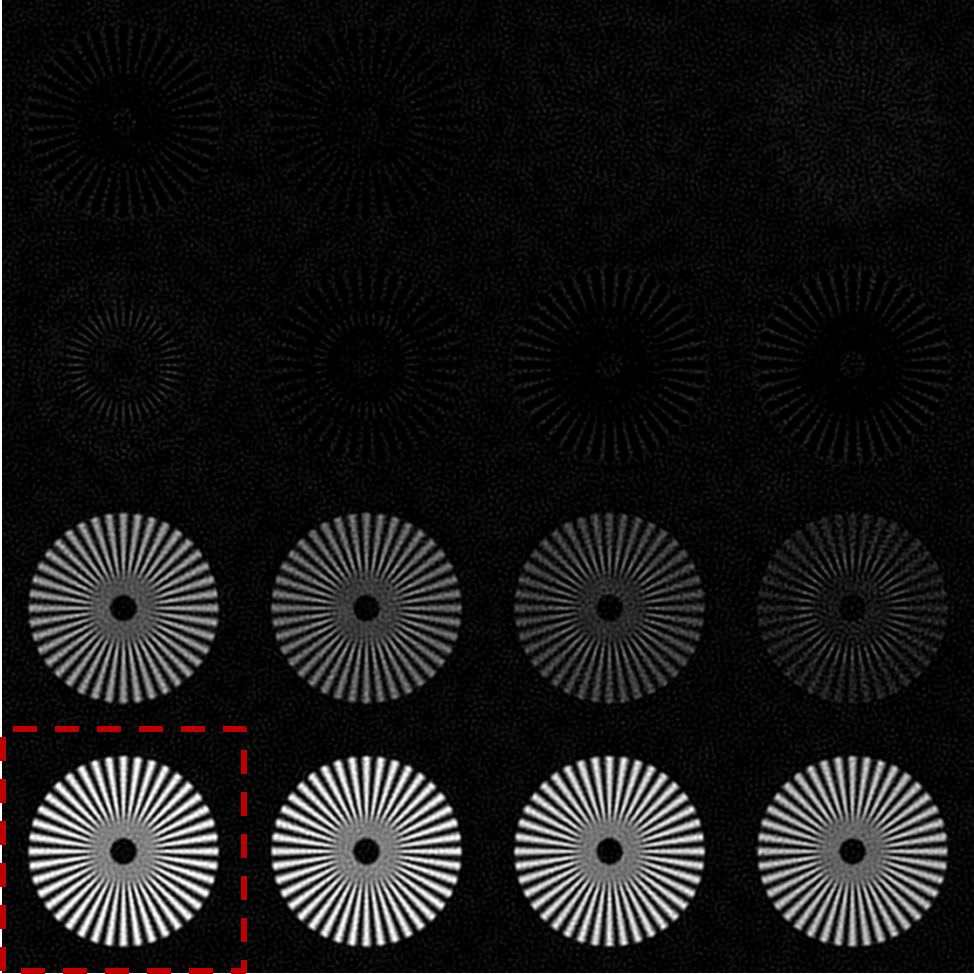}
        \caption{\centering In-focus layer $\hat{\mathbf{o}}_0$, SNR=10}
    \end{subfigure}
    ~
    \begin{subfigure}[b]{0.32\linewidth}
        \centering
        \includegraphics[width=\columnwidth]{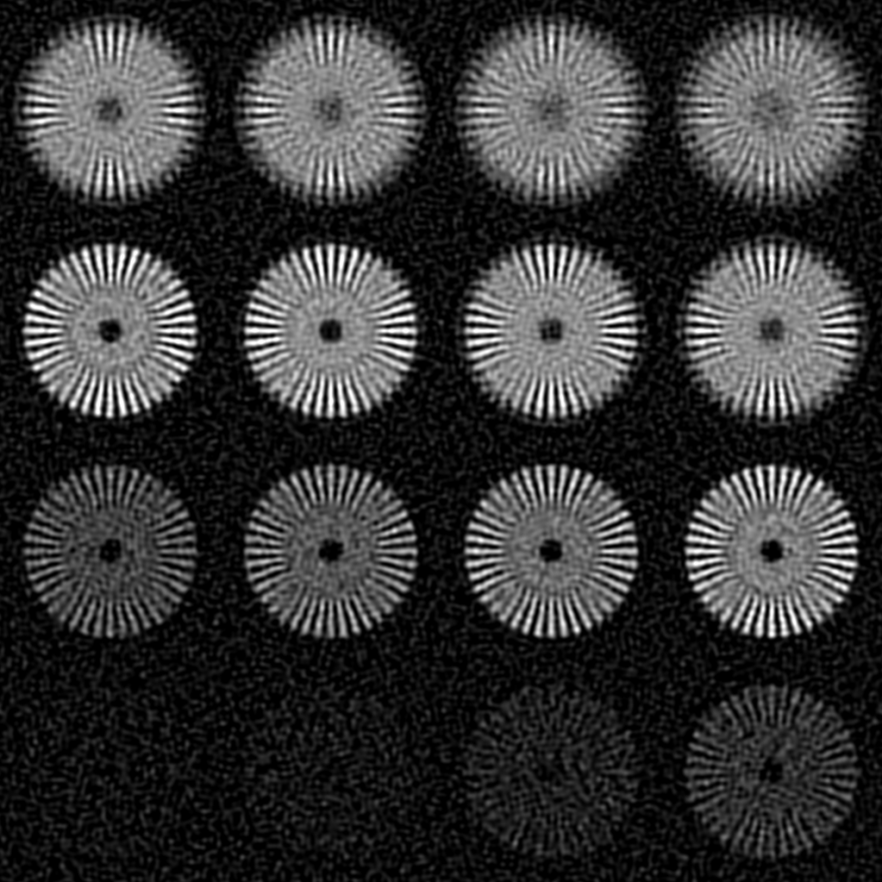}
        \caption{\centering Defocused layer $\hat{\mathbf{o}}_d$, SNR=10}
    \end{subfigure}
    ~
    \begin{subfigure}[b]{0.32\linewidth}
        \centering
        \includegraphics[width=\columnwidth]{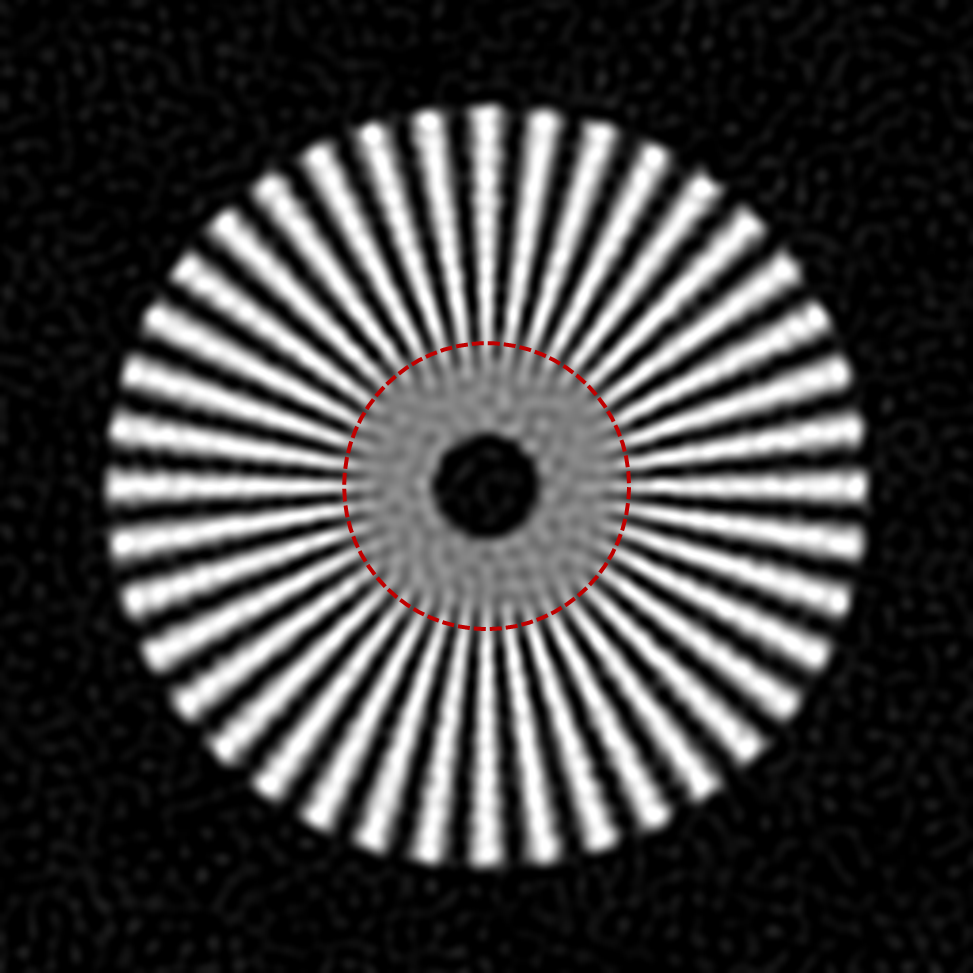}
        \caption{\centering Zoomed-in region, SNR=10}
    \end{subfigure}
    \caption{\small BOSSA-SIM reconstructions of a static object for SNRs of 100 and 10, composed of the in-focus layer $\hat{\mathbf{o}}_0$ (a and d) and the defocused layer $\hat{\mathbf{o}}_{d}$ (b and e). Zoomed views into the red box from $\hat{\mathbf{o}}_0$ are shown (c and f). The zoomed-in region corresponds to the radial target located at object focal plane and the superimposed red circle materializes the resolution limit on conventional widefield image.}
    \label{fig_simu_static}
\end{figure*}

The reconstructions that we obtain with the BOSSA-SIM method for SNRs of 100 and 10 are shown in Fig.~\ref{fig_simu_static}. It clearly shows that the out-of-focus radial targets are rejected in the estimated defocused layer $\hat{\mathbf{o}}_{d}$ (Fig.~\ref{fig_simu_static}, middle column) and we reconstruct in the in-focus layer $\hat{\mathbf{o}}_0$ (Fig.~\ref{fig_simu_static}, left column) essentially the radial targets which are at a distance $u \leq 8$ from the object focal plane. Zoomed view of the reconstructed in-focus layer $\hat{\mathbf{o}}_0$ (Fig.~\ref{fig_simu_static}, right column) indicates that $\hat{\mathbf{o}}_0$ also has an increased lateral resolution compared to the conventional widefield image (Fig.~\ref{fig_simu_data}, right) and the higher the SNR is, the better the achieved resolution is. 
The strength of the OS and the lateral resolution improvement will be quantitatively assessed later. 

\begin{figure}[htbp]
\centering\includegraphics[width=0.55\linewidth]{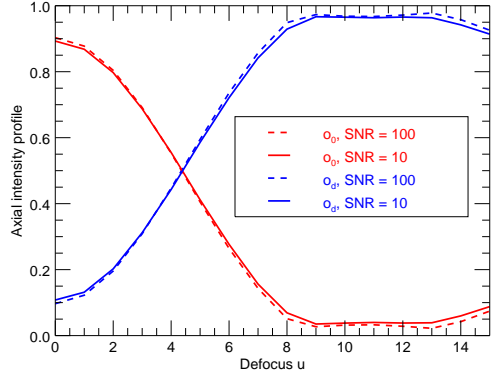}
\caption{\small Distribution of the reconstruction intensity between the reconstructed in-focus layer $\hat{\mathbf{o}}_{0}$ (red plots) and the defocused layer $\hat{\mathbf{o}}_{d}$ (blue plots) for SNRs of 10 (solid line) and 100 (dash line) as a function of the defocus $u$. 
}
\label{fig_static_os}
\end{figure}

Let us firstly analyze how the 16 object layers are allocated between the in-focus and out-of-focus reconstructed layers. We compute the axial intensity profile of both layers as explained in Subsection~\ref{sect_perf_criteria}, which we normalize by the axial intensity profile of the sum of reconstructed layers $\hat{\mathbf{o}}_{0} + \hat{\mathbf{o}}_{d}$ to evaluate how the total reconstruction intensity is distributed among $\hat{\mathbf{o}}_0$ and $\hat{\mathbf{o}}_{d}$. The results are plotted in Fig.~\ref{fig_static_os}.
It shows that, at focal plane $u=0$, $90\%$ of the total intensity is attributed to the in-focus layer which is close to the ground-truth value equal to $87\%$, then the in-focus intensity ratio rapidly decreases with defocus, reaching near-zero values from $u=9$. Inversely, the proportion of intensity allocated to the out-of-focus layer increases with defocus $u$, tending near-one values from $u=9$.
This proves that our reconstruction method is able to correctly distinguish the in-focus and out-of-focus contributions and distribute them among the 2 reconstructed layers $\hat{\mathbf{o}}_{0}$ and $\hat{\mathbf{o}}_{d}$.
The small increase of intensity attributed to $\hat{\mathbf{o}}_{0}$ at $u=15$ is due to a slight rise of the contrast of the projected modulation pattern $C_z$ around this defocus.
Furthermore, we notice that the axial sectioning capability of our reconstructions is robust to noise as it gives almost the same axial intensity distribution for a relative low SNR of 10 and a high SNR of 100.



\begin{figure}[htbp]
\includegraphics[width=0.5\linewidth]{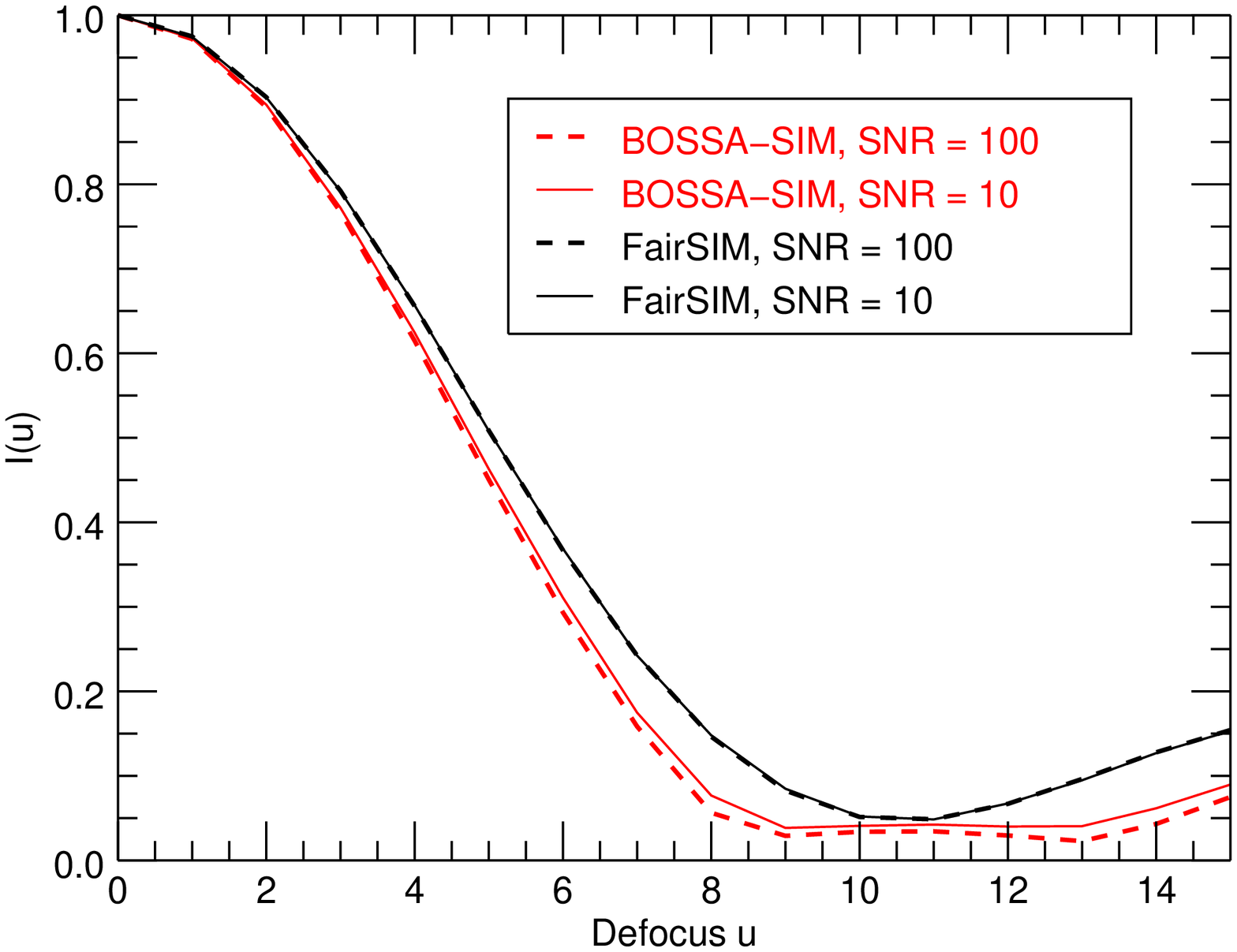}
\includegraphics[width=0.5\linewidth]{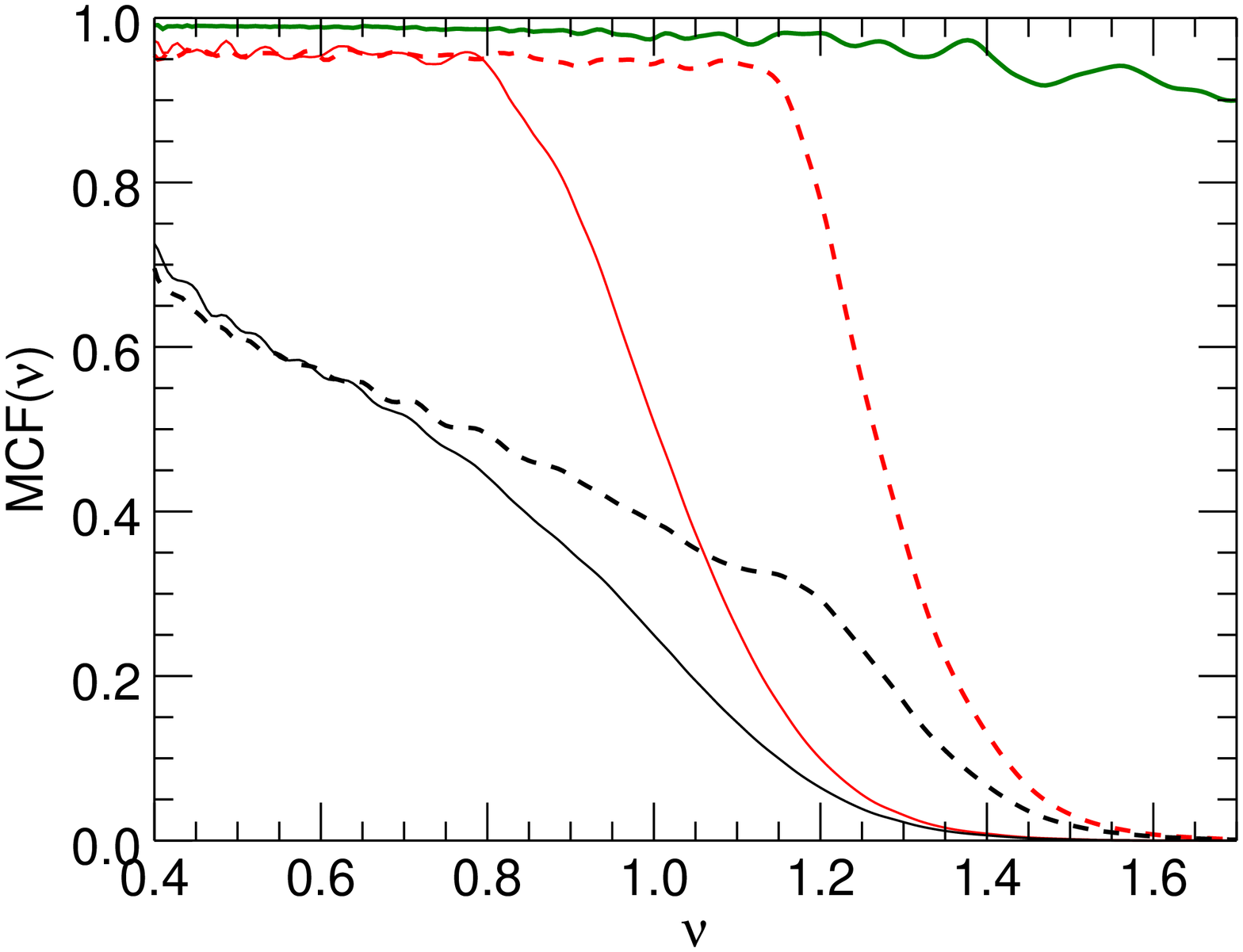}
\caption{\small Performance comparison: Normalized axial intensity profile $I(u)$ as a function of the defocus $u$ (left) and modulation contrast function MCF$(\nu)$ as a function of the normalized spatial frequency $\nu=f/f_c$ (right). $I(u)$ is normalized to unity at zero defocus. Red and black plots show respectively BOSSA-SIM and fairSIM results for SNRs of 100 (dash line) and 10 (solid line). Ground truth MCF is plotted in green.}
\label{figu_static_compare_fairsim}
\end{figure}

The performance of our reconstructions was assessed and compared with that of fairSIM reconstructions for SNRs of 100 and 10. We measured for both methods a normalized reconstruction noise standard deviation of 3\% and 5\% for SNRs of 100 and 10 respectively. We evaluated the achieved OS and SR of the reconstructed in-focus layer $\hat{\mathbf{o}}_{0}$ and the fairSIM reconstructions using the axial intensity profile $I(u)$ and the MCF respectively as explained in Subsection~\ref{sect_perf_criteria}. The results are plotted in Fig.~(\ref{figu_static_compare_fairsim}).
Regarding the OS capability (Fig.~\ref{figu_static_compare_fairsim}, left), we note that BOSSA-SIM provides a slightly better intensity attenuation with defocus than fairSIM for both SNRs, showing that we are able to reject more efficiently the out-of-focus contributions. Furthermore, in both SNRs cases, $I(u)$ is very similar for both methods, showing that the OS is quite robust to noise. We measure a half-width of the axial intensity profile $u_{1/2,BOSSA-SIM}=4.6$ for our method which compares favourably with the value $u_{1/2,fairSIM}=5.0$ for fairSIM. Thus BOSSA-SIM provides a better OS than fairSIM.

Regarding the SR capability (Fig.~\ref{figu_static_compare_fairsim}, right), the contrast of the ground truth radial target is shown (green plot) as a reference and is, as intended, close to 1 for all spatial frequencies. The slight fluctuations are due to sampling and interpolation effects.
As expected, the reconstruction MCFs gradually decrease to 0 when the spatial frequency increases. Both reconstructions achieve SR for both SNRs as their MCF is non null for frequencies higher than the conventional widefield cutoff frequency $\nu=1$. Furthermore, the BOSSA-SIM MCFs are greater than the fairSIM MCFs at every spatial frequency and for both SNR levels, proving that BOSSA-SIM achieves better contrast. 
The measured maximal reconstruction frequency is somewhat higher with our method than with fairSIM: BOSSA-SIM obtains maximal reconstruction frequencies of $1{.}36 f_c$ and $1{.}13 f_c$ for SNRs of 100 and 10 respectively, whereas fairSIM reaches of $1{.}28 f_c$ and $1{.}05 f_c$ for SNRs of 100 and 10 respectively. As expected, the lower the SNR, the lower the maximal reconstruction frequency is, because of the regularization needed to minimize noise amplification.

To sum up, for a given noise level in the reconstruction, BOSSA-SIM achieves more resolved and contrasted reconstructions with a slightly better optical sectioning compared to fairSIM. In addition, the reconstruction parameters are automatically adjusted in BOSSA-SIM whereas they have to be empirically fixed in fairSIM.

\notevf{As a final note, we have performed some additional simulations to explore the behavior of BOSSA-SIM for higher modulation frequencies $f_m$. We found that when $f_m$ increases towards the optical cutoff frequency $f_c$, the gain in lateral resolution increases and simultaneously the OS capability decreases, which is the expected behavior~\cite{wilson_resolution_2011,chetty_structured_2012}. Additionally, when the SNR decreases from 100 to 10, the OS performance decreases noticeably for $f_m \geq 0.8 f_c$ whereas it was almost independent of the SNR for $f_m =0.5f_c$.}



\subsection{Moving object}

\begin{figure}[htbp]
\centering\includegraphics[width=0.47\linewidth]{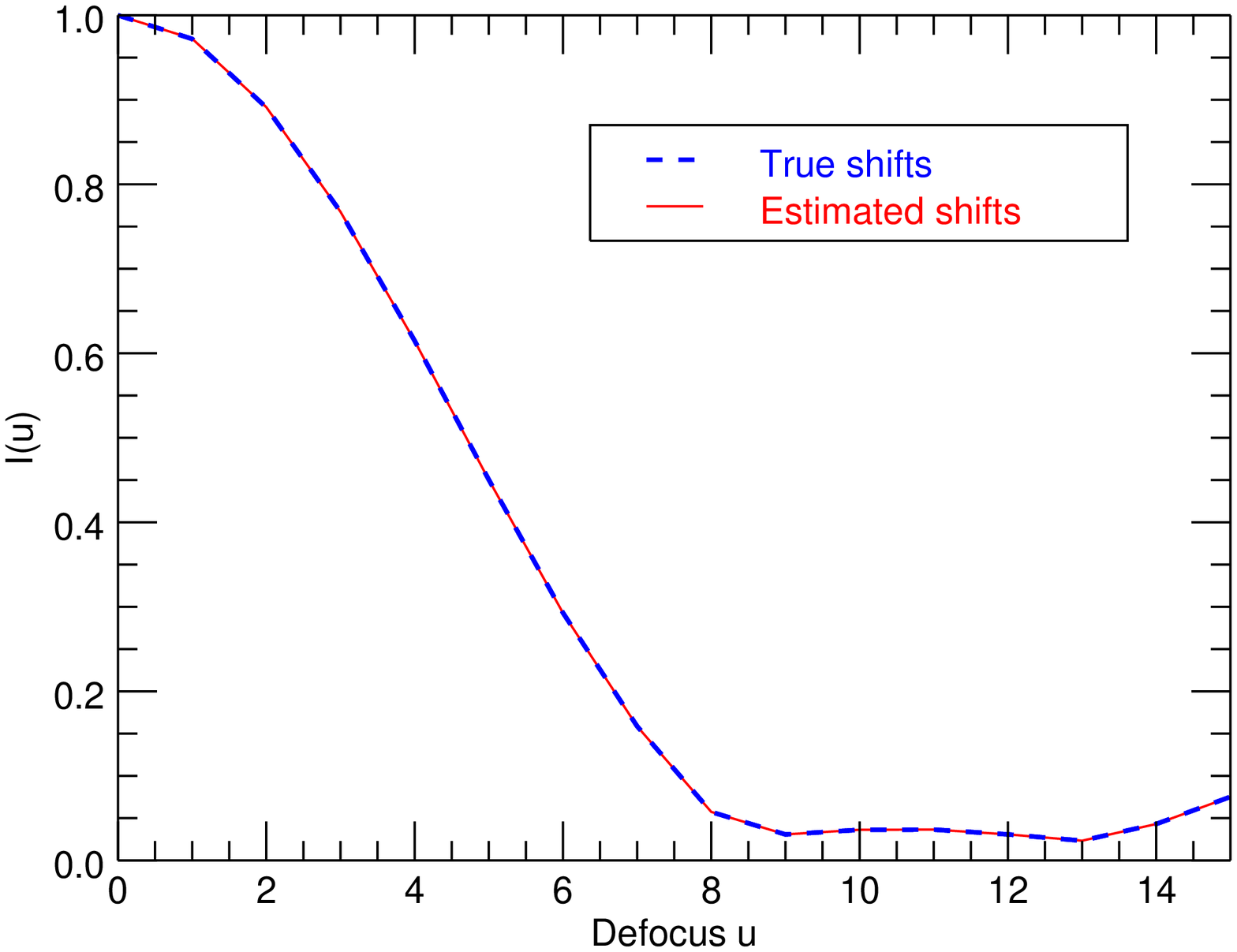}\includegraphics[width=0.47\linewidth]{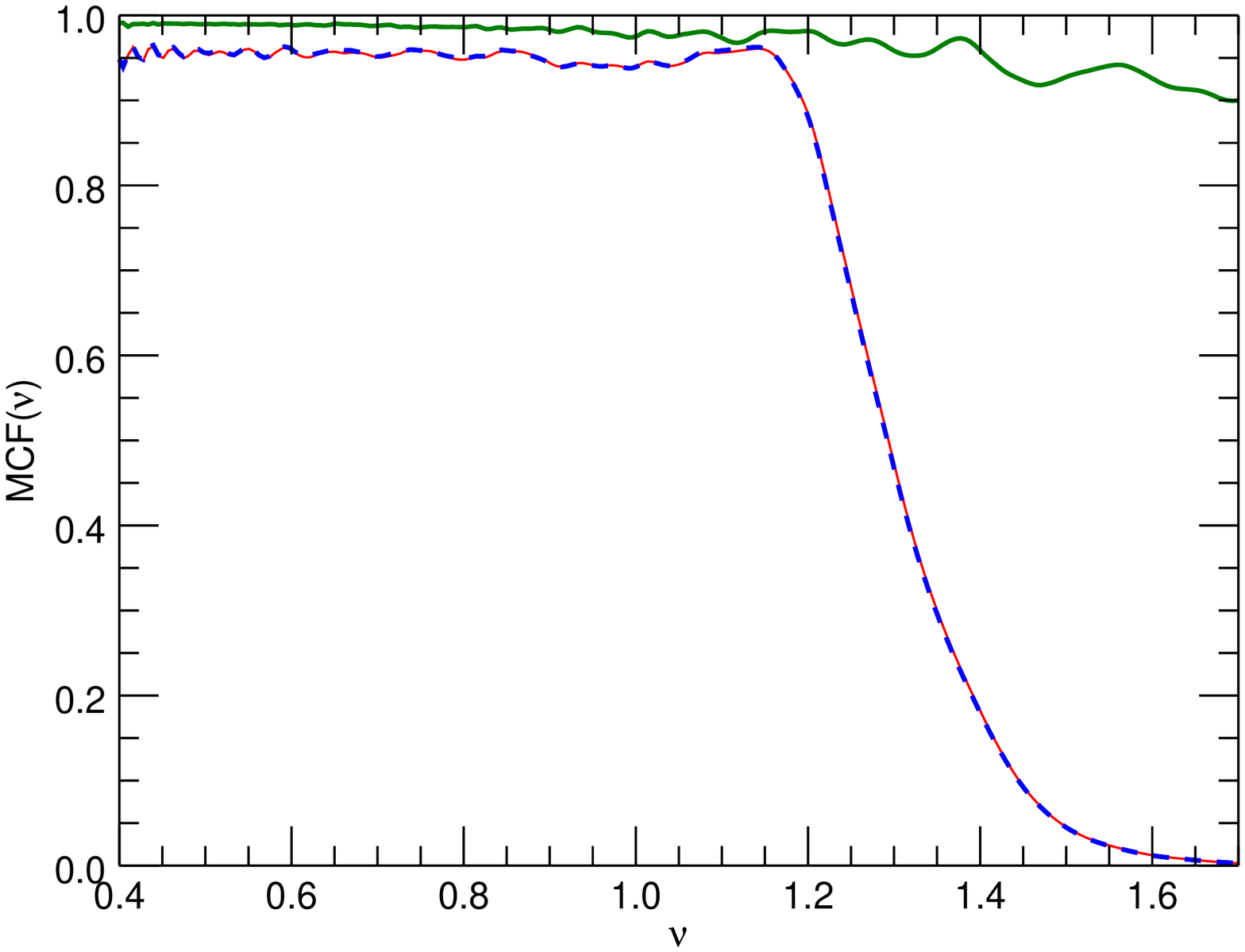}\\[-1em]
\caption{\small Performance assessment: Normalized axial intensity profile $I(u)$ as a function of the defocus $u$ (left) and modulation contrast function $MCF(\nu)$ as a function of the normalized spatial frequency $\nu=f/f_c$ (right). $I(u)$ is normalized to unity at zero defocus. Red and blue plots show BOSSA-SIM results using respectively the estimated and true shifts for a SNR of 10. Green plot shows the ground truth MCF.}
\label{fig_simu_moving_compare}
\end{figure}

In the case of a moving object, we consider simulated SIM data with a SNR of 10.
The main issue when dealing with a moving object compared to a static object, is that the phase of the modulation pattern, i.e. its relative position w.r.t. the object is uncontrolled. We firstly compute BOSSA-SIM reconstruction using the true object shifts to evaluate directly the effect of this uncontrolled phase distribution on the reconstructed object. The axial intensity profile $I(u)$ and the MCF measured are plotted in Fig.~\ref{fig_simu_moving_compare} (blue dash lines). We measure a half-width of the axial intensity profile $u_{1/2}=4.5$ and a maximal reconstruction frequency of $\nu_{max,BOSSA-SIM}=1{,}15$ which are are very similar to the static object case in the same SNR condition (Fig.~\ref{figu_static_compare_fairsim}, red solid lines).

To see the effect of the shift estimation error on the reconstruction, we then perform BOSSA-SIM reconstruction using the shifts estimated with the method described in Subsection~\ref{sect_shift}. The accuracy of our shift estimation was assessed by evaluating the rms shift error in all the data. We found a rms shift errors of $0.018$ pixel which correspond to 0.3 $\%$ of the diffraction size spot. The resulting axial intensity profile $I(u)$ and MCF are superimposed with the plots obtained using the true shifts in Fig.~\ref{fig_simu_moving_compare}, showing that the shift estimation is accurate enough not to reduce the performance of the reconstruction.

Finally, we have showed that our reconstruction method is tailored for moving object, provides both OS and SR, and compares favourably with a state-of-the-art method. 



\section{Reconstruction on experimental microscopy data}\label{sect_experimental}

In order to validate our method on experimental data, we have applied BOSSA-SIM reconstruction on SIM datasets generously provided by the authors of fairSIM~\cite{muller_open-source_2016,fairsim_parameter}.
We have considered the dataset named "OMX LSEC Actin 525nm", consisting of fluorescence images of a stained Actin excited at 488~nm and emitting at 525~nm. The images were acquired with a OMXv4 microscope which uses 3-beam illumination, 3 angles, 5 phases and an objective of numerical aperture $1.4$. Thus each dataset contains 15 images altogether. The illumination patterns projected onto the sample correspond to a 2-order modulation at $f_{m,1}= 0{,}45 f_c$ and $f_{m,2}=0{,}9 f_c$ where $f_c$ is the optical cut-off frequency. It should thus enable both OS and SR.

To apply our method to the given dataset, we firstly calibrated the instrument parameters of our model (Eq.~\ref{eq_model_reduit}) so that they match the characteristics of the OMX microscope. The in-focus PSF $\mathbf{h}_0$ was chosen according to the PSF data provided by fairSIM. It corresponds to an ideal OTF multiplied with an exponential attenuation term of the form $\exp(-a\abs{\nu})$ with $a=0.3$, and $|\nu|=1$ at the optical cutoff frequency. The out-of-focus PSF $\mathbf{h}_{d}$ was obtained from an ideal OTF defocused at $u=8$, and then attenuated with the same exponential attenuation as $\mathbf{h}_0$. 
As for the illumination patterns, the modulation depth of both orders was chosen according the fairSIM data and the modulation frequencies, orientations and phases were estimated from the data. To do so, we identify in Fourier space the peak positions due to modulation in the SIM data with pixel precision, then we filter the data using circular masks of radius 3 pixels centered around the peak positions to essentially keep the contribution due to the illumination pattern, and finally we fit a sinusoidal model on the filtered data using a least squares approach to estimate the illumination pattern parameters (modulation frequencies, orientation and phase).
The reconstruction parameters are then adjusted from the data in the unsupervised way explained in Subsection~\ref{sect_implementation} and the BOSSA-SIM reconstruction were finally performed under the assumption that the sample is static.
As we lack the underlying ground-truth information, we use \notevf{the} fairSIM reconstruction as a reference, with which we check the consistency of our result. The fairSIM reconstruction was performed using parameters provided with the datasets~\cite{fairsim_parameter}. The processed images were apodized in order not to introduce artefacts when computing discrete Fourier transforms and the negative values of the fairSIM reconstruction were thresholded to zero to enhance its displayed contrast.

\begin{figure*}[htbp]
    \centering 
    \begin{subfigure}[b]{0.4\linewidth}
        \centering
        \includegraphics[width=\columnwidth]{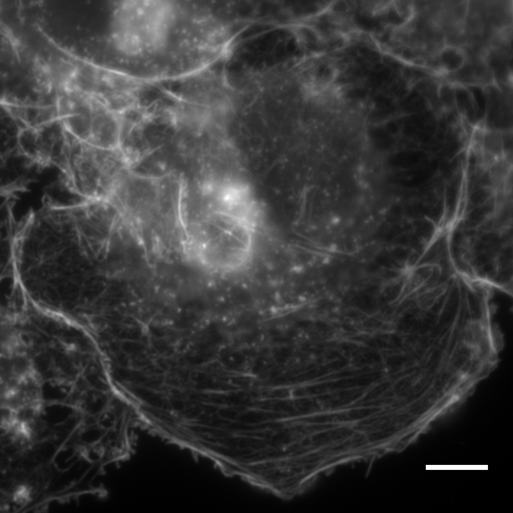}
        \caption{\centering Widefield image}
        \label{subfig:WF_actin}
    \end{subfigure}
    ~
    \begin{subfigure}[b]{.4\linewidth}
        \centering
        \includegraphics[width=\columnwidth]{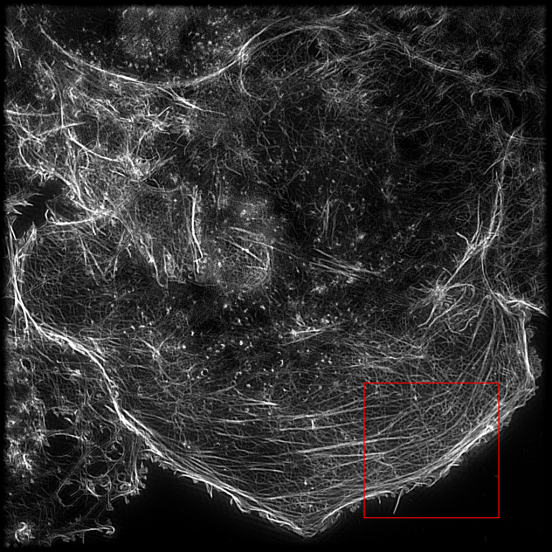}
        \caption{\centering FairSIM reconstruction}
        \label{subfig:fairsim_actin}
    \end{subfigure}
   
    \begin{subfigure}[b]{.4\linewidth}
        \centering
        \includegraphics[width=\columnwidth]{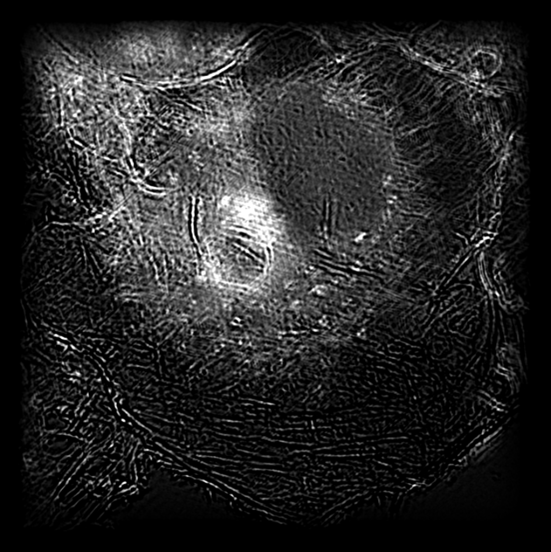}
        \caption{\centering BOSSA-SIM, defocused layer}
        \label{subfig:odefoc_actin}
    \end{subfigure}
    ~
    \begin{subfigure}[b]{.4\linewidth}
        \centering
        \includegraphics[width=\columnwidth]{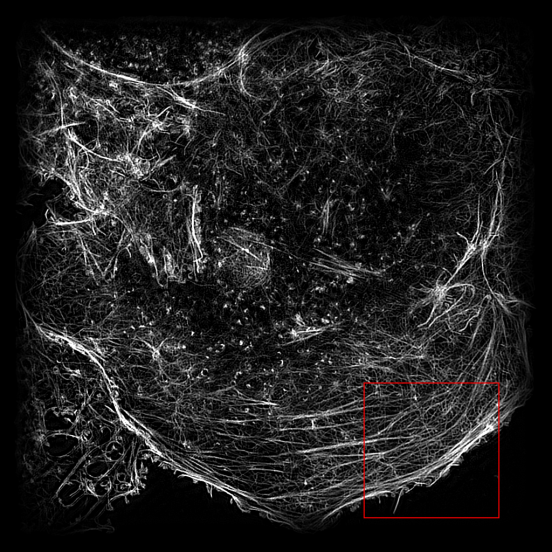}
        \caption{\centering BOSSA-SIM, in-focus layer}
        \label{subfig:o0_actin}
    \end{subfigure}
    \caption{\small Comparison between the widefield image~(a) obtained by averaging the raw data and SIM reconstructions using fairSIM~(b) and BOSSA-SIM~(c and d). Both reconstructions~(b and d) achieve OS and SR compared to the widefield image. Scale bar length:~5 µm. The red box region is zoomed in on Fig.~\ref{fig_exp_zoom}.}
    \label{fig_exp_1}
\end{figure*}

The different reconstructions are shown in Fig.~\ref{fig_exp_1}.
It appears that BOSSA-SIM is able to successfully achieve OS and SR by attributing the out-of-focus contributions such as the blurred circular shape located near the center of the widefield image (Fig.~\myref{fig_exp_1}{subfig:WF_actin}) to the defocused layer (Fig.~\myref{fig_exp_1}{subfig:odefoc_actin}) and by reconstructing in the in-focus layer (Fig.~\myref{fig_exp_1}{subfig:o0_actin}) structures such as filaments better contrasted and resolved compared to the widefield image. Furthermore, the in-focus layer obtained with BOSSA-SIM is consistent in terms of OS and SR with the fairSIM reconstruction (Fig.~\myref{fig_exp_1}{subfig:fairsim_actin}). As we do not know the ground-truth object, we are unable to decide which result is closer to the reality but we can check that our reconstruction method does not introduce any additional artefact compared to fairSIM's.

\begin{figure*}[htbp]
    \centering 
    \begin{subfigure}[b]{0.25\linewidth}
        \centering
        \includegraphics[width=\columnwidth]{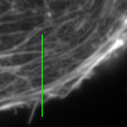}
        \caption{\centering Widefield image}
        \label{subfig:WF_actin_zoom}
    \end{subfigure}
    \begin{subfigure}[b]{.25\linewidth}
        \centering
        \includegraphics[width=\columnwidth]{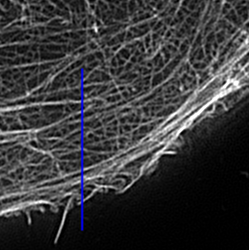}
        \caption{\centering FairSIM}
        \label{subfig:fairsim_actin_zoom}
    \end{subfigure}
    \begin{subfigure}[b]{.25\linewidth}
        \centering
        \includegraphics[width=\columnwidth]{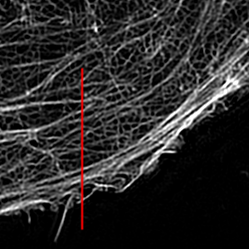}
        \caption{\centering BOSSA-SIM}
        \label{subfig:o0_actin_zoom}
    \end{subfigure}
    \begin{subfigure}[b]{.45\linewidth}
        \centering
        \includegraphics[width=\columnwidth,height=0.8\columnwidth]{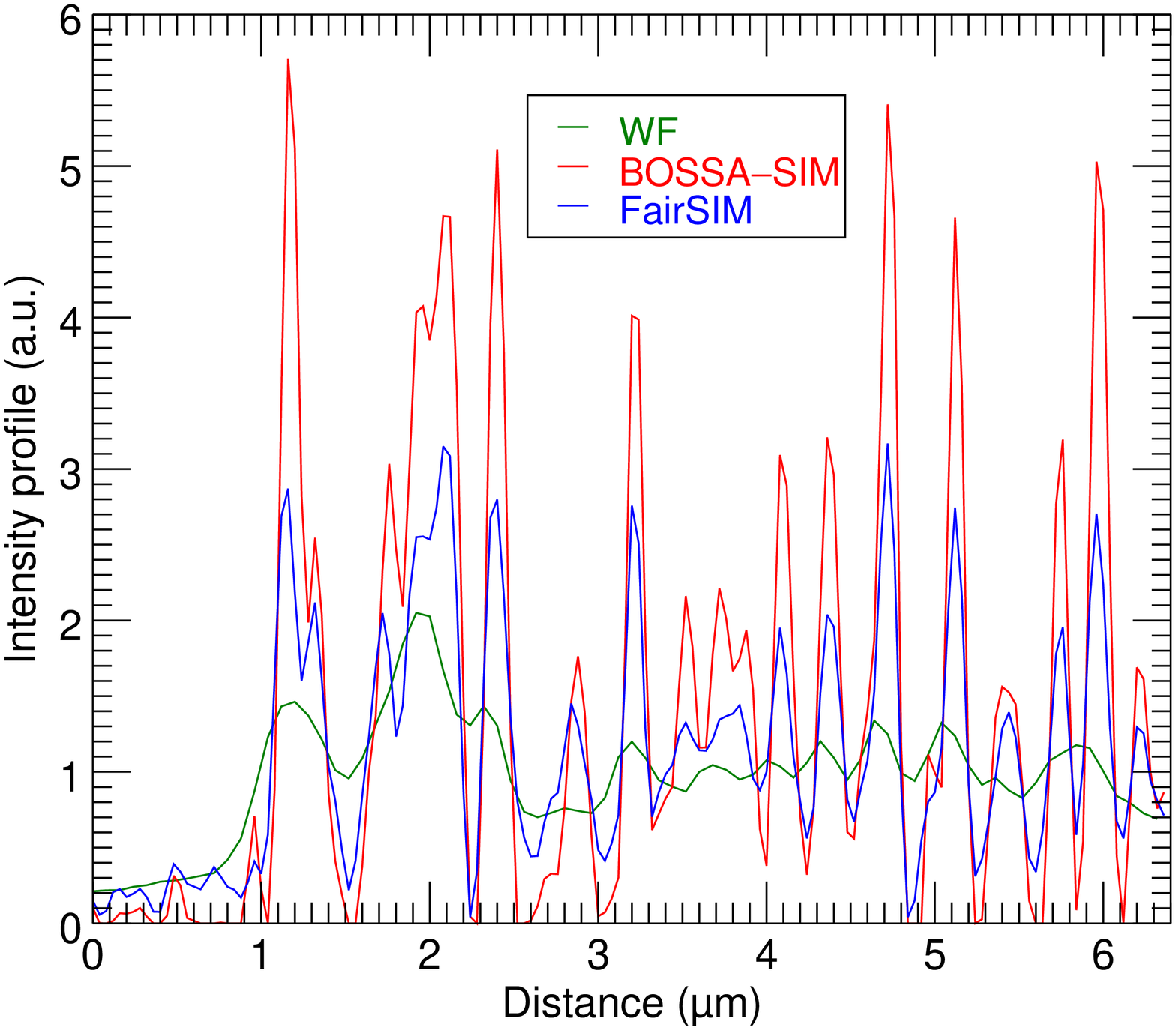}\\[-1em]
        \caption{\centering Intensity profiles}
        \label{subfig:actin_sections}
    \end{subfigure}
    \caption{\small Zoomed-in regions  of area 10x10 µm extracted from the widefield image~(a), fairSIM reconstruction~(b) and BOSSA-SIM's in-focus layer~(c).
    The intensity profiles corresponding to the respectively green, blue and red lines drawn on (a), (b) and (c) are compared in~(d). Each intensity profile has been normalized by its median value and the plot goes from the bottom to the top of the line.}
    \label{fig_exp_zoom}
\end{figure*}

Zoomed-in regions and the plotted normalized intensity sections displayed in Fig.~\ref{fig_exp_zoom} show that the contrast and the capability to resolve tiny structures such as filaments is greatly improved in both SIM reconstructions (Fig.~\myref{fig_exp_zoom}{subfig:fairsim_actin_zoom} and~\myref{fig_exp_zoom}{subfig:o0_actin_zoom}) compared to the conventional widefield image (Fig.~\myref{fig_exp_zoom}{subfig:WF_actin_zoom}). Again, we check that the filament layout reconstructed with BOSSA-SIM is consistent with the one from fairSIM. We also notice on the plots of the normalized intensity profile that BOSSA-SIM reconstructs better contrasted filaments compared to fairSIM, which supports the performance assessment by simulations shown in Fig.~\ref{figu_static_compare_fairsim}.

\section{Conclusion}

We have proposed a novel SIM reconstruction method that achieves both optical sectioning (OS) and super-resolution (SR), in an unsupervised Bayesian framework. Our method is based on a multi-layer imaging model which enables us to properly distinguish the in-focus layer from the out-of-focus contribution while taking into account object motion.
The performance of our reconstruction method, assessed on simulation, exceeds the axial sectioning, contrast and lateral resolution achieved by fairSIM, a state-of-the-art SIM reconstruction technique. When dealing with a moving object, our method still provides OS and SR jointly with a performance that is comparable to the static case. Lastly, we have validated on experimental microscopy data kindly provided by~\cite{muller_open-source_2016} that our reconstruction method yields highly contrasted images with both OS and SR.

\notevf{In order to further improve the performance of the proposed approach, several avenues are worth exploring: (1) the object shifts and the illumination patterns could be jointly estimated with the object using an alternate optimization approach, as done for the illumination patterns in the Blind-SIM method~\cite{jost_optical_2015}; (2) we could study the reconstruction performance with a number of object planes greater than 2, while keeping this number small to minimize the computing time and preserve the robustness to noise; (3) a refinement of the noise model that would take into account the deviation to the Gaussian assumption used here could bring an additional improvement in the reconstructions.}

Finally, we are currently working on an adaptive optics flood-illumination ophthalmoscope with structured illumination capabilities~\cite{gofas-salas_high_2018}, on which we plan to apply our reconstruction method in order to improve both the resolution and the axial sectioning of the imaging system. Such an imaging system would greatly benefit the study of the retinal 3D structure and function in the living eye.

\section*{Funding}
ANR,  RHU \textsc{Light4deaf} (ANR-15-RHUS-0001) project; CNRS, \textsc{Ressources} project of D\'efi Imag’In.

\section*{Acknowledgments}
We thank Frédéric Cassaing, Cyril Petit for fruitful discussions and Antoine Chen, Elena Gofas-Salas and Pedro Mece for initiation of the first author to the retinal imaging bench.




\bibliography{Acronymes,EnglishAcronyms,mybibliography}

\end{document}